\documentclass[aoas,MSNbibl,nameyear,seceqn,dvips]{arximspdf}
\usepackage{mathbh}
\usepackage{dcolumn}
\usepackage{graphicx}

\doi{10.1214/11-AOAS499}
\volume{6}
\issue{1}
\pubyear{2012}
\firstpage{125}
\lastpage{160}

\makeatletter

\newcolumntype{d}[1]{D{.}{.}{#1}}

\newcommand{\Argmax}{\arg\max}

\newcommand{\R}{\mathbb{R}}
\newcommand{\De}{\Delta}
\newcommand{\sig}{\sigma}
\newcommand{\eps}{\varepsilon}
\newcommand{\be}{\beta}
\newcommand{\ga}{\gamma}
\newcommand{\mR}{\mathbb R}

\newcommand{\bss}{\bolds}
\newcommand{\bs}{\mathbf}
\newcommand{\Kapa}{\mathcal{K}}

\makeatother

\begin{document}
\begin{frontmatter}

\title{A stochastic algorithm for probabilistic independent component analysis}
\runtitle{SAEM for probabilistic ICA}

\begin{aug}
\author[A]{\fnms{St\'ephanie} \snm{Allassonni\`ere}\corref{}\ead[label=e1]{Stephanie.Allassonniere@polytechnique.edu}}
\and
\author[B]{\fnms{Laurent} \snm{Younes}\thanksref{T1}\ead[label=e2]{Laurent.Younes@jhu.edu}}
\runauthor{S. Allassonni\`ere and L. Younes}
\affiliation{CMAP, Ecole Polytechnique and CIS, and Johns Hopkins University}
\address[A]{Centre de Math\'ematiques Appliqu\'ees\\
Ecole Polytechnique\\
Route de Saclay\\
91128 Palaiseau\\
France\\
\printead{e1}}
\address[B]{Center for Imaging Science\\
Johns Hopkins University\\
3400 N. Charles Street\\
Baltimore, Maryland 21218\\
USA\\
\printead{e2}}
\end{aug}

\thankstext{T1}{Supported in part by NSF ITR:0427223.}

\received{\smonth{6} \syear{2010}}
\revised{\smonth{7} \syear{2011}}

%
\begin{abstract}
The decomposition of a sample of images on a relevant subspace is a
recurrent problem in many different fields from Computer Vision to
medical image analysis. We propose in this paper a new learning
principle and implementation of the generative decomposition model
generally known as noisy ICA (for independent component analysis) based
on the SAEM algorithm, which is a versatile stochastic approximation of
the standard EM algorithm. We demonstrate the applicability of the
method on a large range of decomposition models and illustrate the
developments with experimental results on various data sets.
\end{abstract}

%

\end{frontmatter}

\section{Introduction}\label{sec1}

Independent Component Analysis (ICA) is a statistical technique that
aims at representing a data set of random vectors as linear
combinations of a fixed family of vectors with statistically
independent coefficients.
It was initially designed to solve source separation problems in
acoustic signals [\citet{bremondmoulinescardoso}] and rapidly
found a
large range of applications, in particular, in medical image analysis
[\citet{Calhoun2001}, \citet{Calhoun2001a}], where ICA has
become one of the
standard approaches. And because it is often valuable to decompose a
large set of variables into simple components, ICA applies more
generally as well [in computer vision
\citet{Bartlett2002}, \citet{bs95},
\citet{Farid2002},
\citet{Liu2003}; and in computational
biology \citet{Liebermeister2002}, \citet{Makeig1997},
\citet{Scholz2004}, etc.].

Often in such problems, the data are high dimensional but have small to
moderate sample size, which complicates statistical analysis.
For example, one challenge in medical imaging is to extract significant
information from spatially \mbox{varying} anatomical or functional signals
drawn from a relatively small number of individuals. A common way to
address this issue is to apply dimension-reduction techniques to reduce
the information to a smaller number of highly informative statistics.
ICA can be used for this purpose, and, in many cases, the
representations it provides are qualitatively very different from those
obtained using decorrelation methods such as principal components
analysis (PCA) [\citet{ufrlSPIEMedia2003}].

ICA can be formulated in terms of a generative model that approximates
the distribution of the data, allowing well-understood statistical
methods to be used for training and validation. ICA represents an
observed $d$-di\-mensional random variable $\bs X$ as
%
%
\begin{equation}
\label{eqICA}
\bs X = \sum_{j=1}^d \beta^j \bs a_{j} ,
\end{equation}
where $(\bs a_1, \ldots, \bs a_d) \in\mR^{d\times d}$ are parameters
(called decomposition vectors) and $\beta^1,\ldots,\beta^d$
are \textit{independent} scalar random variables drawn from a
specified distribution (or family of distributions).
One product of ICA is an estimate of the decomposition matrix
$\bs A = (\bs a_1, \ldots,\bs a_d)$ based on i.i.d. observations $(\bs
X_1,\ldots,\bs X_n)$.
With model (\ref{eqICA}), the independent components
$\beta^1, \ldots, \beta^d$ can be computed from $\bs X$ by
inverting $\bs A$.
A variety of methods and criteria have been proposed to estimate
either $\bs A$ or $\bs W = \bs A^{-1}$ (see
\texttt{\href{http://www.tsi.enst.fr/icacentral/index.html}{http://www.tsi.}
\href{http://www.tsi.enst.fr/icacentral/index.html}{enst.fr/icacentral/index.html}}, from which
some algorithms may be accessed).
For example, in \citet{yeredor2001}, $\bs A$ is seen as a~joint
diagonalizer of a~set of estimated correlation matrices. In
\citet{BellSejnowsky} and \citet{ekk2000}, standard estimation
procedures, like maximum entropy or minimum Kullback--Leibler
divergence, are used with specified distributions for the independent
components.

We will also use a model-based formulation in this paper,
but it is important to mention that a large class of algorithms have
also been defined for distribution-free representations (based on the
so-called negentropy---non-Gaussian entropy---and cumulant
expansion), including the widely used FastICA method
[\citet{hoNeuralComputation97}], as well as algorithms proposed in
\citet{Learned-miller03} or \citet{bj03}, which maximize the
independence (with respect to some criteria) of the components using
the semi-parametric model of ICA. A comparison study has been
made in \citet{cardoso1999}, using high-order measures to assess
component independence.

One of the drawbacks of ICA is that it does not come (like PCA does)
with a well-defined method to select the most important components. In
the original formulation, the number of independent components is
equal to the dimension of the variables, so that the decomposition
is achieved without dimensional reduction. This leads to computational
and overfitting issues when dealing with high-dimensional data and
small sample sizes, and a
lack of interpretability of the obtained results.\vadjust{\goodbreak}

Probabilistic ICA
(alternatively called noisy ICA, or independent factor analysis,
although we will reserve the latter term to a more specific method in
which factors are Gaussian mixtures) assumes a small number of
independent components, with a residual term which is modeled as
Gaussian noise. The explicit model is therefore given by
%
%
\begin{equation}
\label{eqNoisyICA}
\bs X = \sum_{j=1}^p \beta^j \bs a_{j} + \sigma\bss\eps,
\end{equation}
where $(\bs a_1, \ldots, \bs a_p)\in\mR^{d\times p}$ now represent
$d\times p$ parameters (to be compared to the $d\times d$ matrix,
$\mathbf A$, estimated in the standard ICA model), $\beta^1, \ldots,
\beta^p$ are independent scalar random variables and $\bss\varepsilon
$, the
noise, follows a standard normal distribution (we will take the
standard deviation, $\sig$, to be a fixed scalar, also a
parameter). Such models therefore represent the $d$-dimensional input
vector, $\bs X$, by $p$ scalar components, achieving the required
dimensional reduction.

The ICA training algorithms (e.g., estimating $\bs W$) do not
generalize to probabilistic ICA. In particular, the $d$-dimensional
vector $\bs X$ is modeled as a~function of the $(p+d)$-dimensional
variable $(\bss\beta, \bss\varepsilon)$ and we have partial
observations. A~possible approach is to first implement some dimension
reduction to the data, typically projecting $\bs X$ on the $p$ first
principal components to eliminate the residual, before applying
standard ICA to the projection [\citet{ccoaICMLA2008},
\citet{vsptArXiv09}]. But this procedure does not necessarily
retrieve the model described in~(\ref{eqNoisyICA}) (especially when the
noise has a~large variance), and training probabilistic ICA in a way
which is consistent with this statistical model certainly is a more
satisfactory approach.

The numerical method described in this paper estimates the maximum
likelihood estimator associated to (\ref{eqNoisyICA}), where the
likelihood is for the observations, $\bs X$, therefore averaging over
the unobserved components $\bss\beta$. This differs from the solution
which is
often adopted in the literature, which consists in maximizing the joint
likelihood of $\bs X$ and $\bss{\beta}$,
simultaneously in the parameters and in the unobserved variables
[\citet{hyvarinen}]. This latter method attempts to solve the
parametric estimation and
hidden variable reconstruction problems at the same time.
However, the estimation of both~$\bs X$ and $\bss{\beta}$ is not always
a good choice, because it can lead to biased estimators: as we will
show in our experiments, these approaches
have good results when the noise level
is small [as already noticed in \citet{lpIWICABSS2000}], but these
results can significantly degrade otherwise [see
Section \ref{secExp}, or \citet{AAT} for a similar observation
made in
a different context]. In contrast, averaging over the unobserved
variables takes the whole distribution into account, which becomes
important as soon as the posterior distribution is not unimodal, with
its mean equal to its mode. The reconstruction problem (estimating $
\bss{\beta}$
from $\bs X$), which is also important, for example, to define
efficient lossy
compression methods, can be solved \textit{afterward} using the estimated
parameters. Estimation and reconstruction are, in this regard, two
separate problems.

When independent components are modeled as mixtures of Gaussians, as
done in \citet{mcg97} for blind source separation and blind
deconvolution, or with independent factor analysis (IFA), as
introduced in \citet{Attias99}, maximizing the likelihood of the
observations (averaging over the nonobserved independent components)
can be done using the expectation--maximization (EM) algorithm. In this
particular case, this algorithm can be derived with closed form
formulae and explicit computations. But, even in this special case
(mixture of Gaussians), the EM algorithm can become computationally
prohibitive, especially when the number of components is large.
For general component distributions, the explicit evaluation of conditional
expectations given observations constitutes an infeasible task, and
only Markov chain Monte Carlo (MCMC) approximations remain
available. Replacing explicit formulae by Monte Carlo approximations
in the $E$-step of the EM algorithm leads to the MCEM algorithm,
introduced in \citet{mcg97}. MCEM, however, still is a highly
computational procedure, with many Monte Carlo samples required at
each update of the parameters.

We suggest using an alternative approach for maximum likelihood
estimation, relying on a stochastic approximation to
the EM algorithm [called SAEM, \citet{DLM}] which only requires being
able to
sample from this conditional distribution.
Instead of running a long Monte Carlo simulation at each $E$-step, as
MCEM does, this algorithm interlaces sampling with the \mbox{$M$-step},
requiring only a single new sample between two parameter updates.
This algorithm compensates the larger convergence time (in number of steps)
generally associated to stochastic approximations by much simpler
iteration steps.
This algorithm has been proposed and proved convergent under some weak
conditions in \citet{aktdefmod}. A comparison between the MCEM and SAEM
is proposed as part of the
experiments provided here.

Another advantage of our learning algorithm is that it applies to many
different probabilistic distributions. There are almost no restrictions
to the range of statistical models that can be used for the unobserved
independent variables. As examples, we will present in this paper
different models that all fit into this same framework, but which
correspond to different statistical contexts. They will be introduced in
Section \ref{sec2}. The parametric
estimation method, including the SAEM algorithm, is described in
Section \ref{secMAP} and the reconstruction of hidden variables is
discussed in Section \ref{secreconstruction}. Experimental results
with both synthetic and real data are presented in Section
\ref{secExp} where we also provide some comparison with the EM (when
feasible), MCEM and FastICA, three of the most used algorithms.

\section{Models}
\label{sec2}

We start with some general assumptions on the data, that will be made
specific in the experiments.
We assume that the observation is a~set of vectors which take
values in $\mR^d$. Let
$\bs X_1,\ldots,\bs X_n$ be the training observations, which are
assumed to
be independent and identically distributed. We will denote by $\bs X$ a
generic variable having the same distribution as the~$\bs X_k$'s. The $j$th
coordinate of $\bs X$ (resp., $\bs X_k$) will be denoted $X^j$
(resp.,~$X_k^j$).\looseness=-1

We assume that $\bs X$ can be generated in the form
%
%
\begin{equation}
\label{eqobslaw}
\bs X = \bss\mu_0 + \sum_{j=1}^p \beta^j \bs a_j +
\sigma\bss{\eps} ,
\end{equation}
where $\bss{\mu}_0\in\mathbb R^d$, $\bs a_j \in\mR^d$ for all
$j\in\{
1,\ldots, p \}$, $\bss{\varepsilon}$ is a standard $d$-dimensional
Gaussian variable and $\beta^1, \ldots, \beta^p$ are $p$
independent scalar variables, the distribution of which being specified
later. Let $\bss{\beta}$ denote the $p$-dimensional variable $\bss
{\beta}=
(\beta^1, \ldots, \beta^p)$. To each observation $\bs X_k$ is therefore
associated hidden realizations of $\bss{\beta}$ and $\bss{\eps}$,
which will
be denoted $\bss{\beta}_k$ and $\bss{\eps}_k$.

Denote $\bs A = (\bs a_1, \ldots, \bs a_p)$. It is a $d$ by
$p$ matrix and one of the parameters of the model. Another parameter is
$\sigma$, which will be a scalar in our case (a diagonal
matrix being also possible). Additional parameters will appear in
specific models
of $\bss\beta$ which are described in the following subsections. In
some of these
models, it will be convenient to build $\bss\be$ as a function of new
hidden variables, which will be denoted $\bs Z$.

The models that we describe are all identifiable, with the obvious
restriction that $\bs A$ is
identifiable up to a permutation and a sign change of its columns (the
latter restriction being needed only when
the distribution of~$\bss\be$ is symmetrical). This fact derives from
identifiability theorems for factor analysis, like Theorem 10.3.1 in
\citet{klrMathsStat}.

\subsection{Logistic distribution (Log-ICA)}

We start with one of the most popular models, in which each $\beta^j$
follows a logistic
distribution with fixed parameter ${1}/{2}$. The associated cumulative
distribution function is
$P(\beta^j\leq t) = 1/(1 + \exp(-2t))$.

For this model, the parameters to
estimate are $\theta= (\bs A,\sigma^2,\bss\mu_0)$. Hidden variables
are $\bs Z
= \bss\be$ and $\bss\eps$. This is the model introduced in the original
paper of Bell and Sejnowsky [\citet{bs95}], and probably one of the
most commonly used parametric models for ICA. One reason for this is
that the logistic probability density function (p.d.f.) is easy to
describe, smooth, with a shape
similar to the Gaussian, but with heavier, exponential, tails. Note
that, for identifiability reasons,
one cannot use Gaussian distributions for the components.

\subsection{Laplacian distribution (Lap-ICA)}
A simple variant is to take $\be^j$ to be Laplacian with density
$e^{-|t|}/2$. The parameter still is $\theta= (\bs A,\sigma^2,
\bss\mu_0)$. Hidden variables are $\bs Z = \bss\be$ and\vadjust{\goodbreak}
$\bss\eps$.

The resulting model is very similar to the previous one with similar
exponential tails, with the noticeable difference that the Laplacian
p.d.f. it is not differentiable in $0$. One consequence of this is that
it leads to sparse maximum a posteriori reconstruction of the hidden
variables (cf. Section~\ref{secreconstruction}).\looseness=1

\subsection{Exponentially scaled Gaussian ICA (EG-ICA)}
\label{secmod3}
In this model, we let $\beta^j = s^j Y^j$ where
$\bs Y$ is a standard Gaussian vector, 
$s^1,\ldots, s^p$ are independent exponential
random variables with parameter 1, also independent from $Y$ and
$\bss\eps$. In this case, we can write
%
%
\begin{equation}
\label{eqmod3}
\bs X = \bss\mu_0 + \sum_{j=1}^p s^jY^j\bs a_j + \bss\sig
\varepsilon.
\end{equation}
Hidden variables are $\bs Z = (\bs s, \bs Y)$ and $\bss\eps$,
and the parameter is $\theta= (\bs A, \sig^2, \bss\mu_0)$.

The p.d.f. of $\be= sY$ is given by $g(\be) = \int_0^\infty\exp
( -\frac12 y^2 -\frac{\be}{y}
) \,\frac{dy}y$. It tends to infinity at $\be= 0$, and has
subexponential tails, because $\log[P(\be^i > t)]$ is asymptotically
proportional to $(-t^{2/3})$ (see the \hyperref[app]{Appendix} for
details). It
therefore allows for higher sparsity and more frequent large values of
the component coefficients. This may help to overcome the variability
in intensity which appears in medical images for examples. If we think
in terms of source separation, the source has its own intensity and
observations may require a large range of intensity around this ``mean.''
It is also important to notice that, in spite of its increased
complexity, this model can be implemented and learned as simply as the
previous two using the algorithm that is proposed here.

\subsection{Independent factor analysis (IFA)}

The IFA [\citet{Attias99}, \citet{mmkCUP2001},
\citet{mcg97}] model is a~special case of probabilistic ICA in
which the distribution of each coordinate~$\beta ^j$ is assumed to be a
mixture of Gaussians. We will here use a~restricted definition of the
IFA model which will be consistent with the other distributions that we
are considering in this paper, ensuring that the~$\beta^j$'s are
independent with identical distribution, and that this distribution is
symmetrical.

More precisely, we will introduce two new sets of hidden variables, the
first one, denoted $(t^1, \ldots, t^p)$, represents
the class in the mixture model,
and the second one, denoted $(b^1, \ldots, b^p)$, is a random sign
change for
each component. Each
$t^j$ takes values in the finite set $\{0, 1, \ldots, K\}$, with
respective probabilities $w_0, \ldots, w_K$, and $b^j$ takes values
$\pm1$ with probability $\frac12$. We then
let
\[
\beta^j = b^j \sum_{k=1}^p m_k \delta_k(t^j) + Y^j,
\]
where $Y^j$ is standard Gaussian. In other terms, $\beta^j$ is a
mixture of $2K+1$ Gaussians with unit variance, the first one being
centered, and the following ones having means $m_1, -m_1, m_2, -m_2,
\ldots.$

The parameters of this model are therefore $\theta=(A,\sigma^2,
(w_k,m_k)_{1\leq k\leq K})$. Hidden variables are $\bs Z = (\bss\be,
\bs b, \bs t)$.
Note that, even if we use a simplified and symmetrized version of the
model originally
presented in \citet{Attias99}, the stochastic approximation
learning algorithm that will be designed in Section~\ref{secalgo}
immediately extends to the general case where the means depend on the
index $j$.

\subsection{Bernoulli-censored Gaussian (BG-ICA)}

\label{secbg}

In contrast with the logistic or Laplacian models for which
coefficients vanish with probability zero, we now introduce a discrete
switch which ``turns them off'' with positive probability.
Here, we model the hidden variables as a Gaussian-distributed scale
factor multiplied by a Bernoulli random
variable.
We therefore define
$\beta^j = b^j Y^j$, using the same definition for $\bs Y$ as in
Section \ref{secmod3} and letting
$b^j$ have a Bernoulli distribution with parameter $\alpha=
P(b^j=1)$. We assume that all
variables $b^1, \ldots, b^p, Y^1, \ldots, Y^p, \eps$
are independent.
The complete model for~$\bs X$ has the same structure as before, namely,
%
%
\begin{equation}
\label{eqmod4}
\bs X = \bss\mu_0 + \sum_{j=1}^p b^jY^j\bs a_j + \sig
\bss\eps.
\end{equation}

Parameters in this case are $\theta= (\bs A, \sig^2, \alpha,
\bss\mu_0)$ and hidden variables are $\bs Z = (\bs b, \bs Y)$ and
$\bss\eps$.

Using a censoring distribution in the decomposition is a very simple
way to enforce sparsity in the resulting model. The population is
characterized by a set of $p$ vectors, however, each subject is only
described by a subset of these $p$ vectors corresponding to the active
ones. The probability of the activation of the vectors is given by
$\alpha$. As $\alpha$ increases, the sparsity in the subject
decomposition increases as well, whereas the dimension to explain the
whole training set may remain equal to~$p$. Censored models therefore
arise naturally in situations where independent components are not
expected to always contribute to the observed signals. This often
occurs in spatial statistics, in situations for which observations
combine basic components in space, not necessarily occurring all
together. We will see an example of such a situation with handwritten
digits where components can be interpreted as common parts of some of
the digits, but not all, and therefore should not be selected every
time. Functional magnetic resonance images (fMRIs), for which ICA
methods have been extensively used [\citet{Calhoun2001},
\citet {Calhoun2001a},
\citet{Makeig1997}], are also important
examples of similar situations. These three-dimensional images indicate
active areas in the brain when a subject executes a specific cognitive
task. People generally interpret components as basic processing units
that interact in a complex task, but these units are not expected to be
involved in every task for every subject. Similarly, genomic data,
where a gene can activate a protein or not for particular patients, may
fall into this context as well.

We now describe some possible variants within the class of censored models.

\subsection{Exponentially scaled Bernoulli-censored Gaussian (EBG-ICA)}

Combining EG- and BG-ICA, so that a scale factor and a censoring
variable intervene together, we get a new
complete model for $\bs X$ given by
%
%
\begin{equation}
\label{eqmod5}
\bs X = \bss\mu_0 + \sum_{j=1}^p s^j b^jY^j\bs a_j + \sig
\bss\eps.
\end{equation}
Since the exponential law has fixed variance, the parameters of
interest are the same as in the BG-ICA model, that is, $\theta= (\bs A,
\sig^2, \alpha, \bss\mu_0)$. The hidden variables are $\bs Z = (\bs
s, \bs b, \bs Y)$ and $\bss\eps$.

\subsection{Exponentially-scaled ternary distribution (ET-ICA)}

The previous models include a switch which controls whether the
component is present in the observation or not. One may want to further
qualify this effect as ``activating'' or ``inhibiting,'' which can be
done by introducing\vspace*{1pt} a discrete model for $\bs Y$, each
component taking values $-1$, $ 0$ or $1$. We define $\beta^j = s^j
Y^j$, where $s^1, \ldots, s^p$ are i.i.d. exponential variables with
parameter 1. We let $\ga= P(Y^j=-1) = P(Y^j=1)$, providing a symmetric
distribution for the components of $\bs Y$. As before, all hidden
variables are assumed to be independent. The model is
%
%
\begin{equation}
\label{eqmod6}
\bs X = \bss\mu_0 + \sum_{j=1}^p s^jY^j\bs a_j + \sig
\bss\eps.
\end{equation}

Hidden variables here are $\bs Z = (\bs s, \bs Y)$ and
$\bss\eps$, the
parameter being $ \theta= ( \bs A , \sigma^2 , \gamma, \bss\mu_0 )$.

The interpretation of the decomposition is that each component has
a~fixed effect, up to scale, which can be positive, negative or
null. The model can therefore be seen as a variation of the
Bernoulli--Gaussian where the effect can be a weighted inhibitor as
well as a weighted activator.
This allows selective appearance of decomposition vectors and
therefore refines the characterization of the population.

This particular model makes all its sense when trying to model the
generation of data with nonzero mean. Going back to our fMRI example,
the mean image is more likely to be an active brain since all the
patients are subject to the same cognitive task and the activation is
always positive or zero. This will create some active\vadjust{\goodbreak} areas in the
mean brain ($\bss\mu_0$). However, as we already noticed, these areas
can be active or not depending on the subject participating to the
experiment. This can be modeled by a weighted activation or
inhibition of its areas around the mean through the corresponding
decomposition vectors. The decomposition vectors are still expected to
correspond to the different active zones. This is what this model
tries to capture. We will see in the experiments that it also applies
to the handwritten digits.

\subsection{Single-scale ternary distribution (TE-ICA)}

The previous model can be simplified by assuming that
the exponential scale factor is shared by all the components, that is,
we let $\beta^j = s Y^j$, where $s$
is exponential with parameter 1, and $Y^j$ has the same ternary
distribution as in the ET-ICA model. The decomposition now is
%
%
\begin{equation}
\label{eqmod7}
\bs X = \bss\mu_0 + s\sum_{j=1}^p Y^j \bs a_j + \sig
\varepsilon.
\end{equation}

Hidden variables here are $(s, \bs Y)$, the
parameter being $ \theta= ( \bs A , \sigma^2 , \gamma,\bss\mu_0
)$. Notice that
this model is not explicitly an ICA decomposition, since the
components are only independent given the scale factor. Notice also
that we assume
that the scaling effect acts on the components, not on the observation
noise which remains unchanged.

Probabilistic-ICA in general is obviously a very efficient
representation for lossy
compression of random variables, since, if the noise is neglected, and
as soon as the parameters $\bss\mu_0$
and $A$ are known, one
only needs to know the realization of $\bss\be$ (hopefully with
$p\ll d$)
to reconstruct an approximation to the signal. In the present model,
the transmission of $\bss\be$
only requires sending the scalar scale factor, $s$, and $p$ ternary
variables. If many components vanish (i.e., if $\ga$ is significantly
smaller than $1/2$), compression is even more efficient.

In this model (and for the previous two also), the sparsity of the
representation will obviously depend on the number of selected
components, $p$, that we suppose given here. When $p$ is too small, it
is likely that the model will find that censoring does not help and
take $\ga=1/2$ (or $\alpha= 1$ in the Bernoulli--Gaussian model).
Adding more components in the model generally results in $\alpha$ and
$\ga$ decreasing, enabling some components to be switched off. This
effect is illustrated in Section~\ref{secExp}.

Finally, let's remark that, although the results in
\citet{klrMathsStat} do not directly apply to this model (the
components are not independent, since they share the same scale
factor), they can be applied to the conditional distribution given
the scale to prove identifiability (since the scale factor distribution
is fixed).

\subsection{Playing with the average}
\label{secmean}

Clearly, all the previous models admit a~centered submodel in which
$\bss\mu_0 = 0$, which might be preferred in some cases. In this case
($\bss\mu_0=0$), it may be interesting to allow for some shift in the
distribution\vadjust{\goodbreak} of the components, replacing $\be^j$ by $\mu+ \be^j$
where $\mu$ is a one-dimensional parameter. This is therefore
equivalent to modeling $\bss\mu_0 = \bs A \bss\mu$ where $\mu$ is a
$p$-dimensional vector with all coordinates equal to $\mu$. When
dealing with scaled, or censored models, one can decide to apply the
shift before or after censoring or scaling. For example, one can
define a shifted Bernoulli--Gaussian model by replacing $Y^i$ by $\mu+
Y^j$ in Section \ref{secbg}, which results in shifting $\be^j$ only
when it is not censored.

Another choice that can also be
interesting is to model the signal with a random, scalar,
offset (or AC component). One way to achieve this is to impose that one
of the columns of
the matrix $\bs A$ is the $d$-dimensional vector $(1, \ldots,
1)^T$. In this case, it is
natural to separate the distribution of the offset coefficient from the ones
of other components, as customary in compression (the offset
coefficient should not be censored, e.g.). A
simple choice is to provide it with a logistic or Laplacian
distribution. This is illustrated in the next model.

\subsection{Single-scale ternary distribution with offset (TEoff-ICA)}

In this model the mean $\bss\mu_0$ is not a parameter and is not the
same for all the observed vectors, so that this random effect (in
opposition to the fixed effect it had in the previous models) now is
a hidden variable. We furthermore assume that this random variable,
denoted $\bss\mu$, takes the form $\bss\mu=(\mu,\ldots,\mu)\in\R
^d$ where~$\mu$ is Laplacian. So $\mu$ can be interpreted as an offset acting
simultaneously on all coordinates of $\mathbf X$.
This yields the following model:
%
%
\begin{equation}
\label{eqTEoff}
\bs X = \bss\mu+ s\sum_{j=1}^p Y^j \bs a_j + \sig\eps,
\end{equation}
where $s$ follows an exponential distribution with parameter 1 and
$Y^j$ are ternary variables with $\ga= P(Y^j=-1) =
P(Y^j=1)$. The hidden variables are $(s, \bs Y, \mu)$ and the
parameters $(A,\sigma^2 , \gamma)$.

Introducing observation-dependent offset and scale effects is useful
when dealing with uncalibrated observations. This is typical, for
example, with micro-array data, for which strong variations in
calibration can occur among different patients. This is also common for
signal and image processing, for which interpretation often needs to be
performed in a way which is invariant, or robust, to offset or scale effects.

\section{Maximum likelihood estimation}
\label{secMAP}
\subsection{Notation}

The previous models are all built using simple generative relations
$\bs Z\to\bss\be$ and $(\bss\be, \bss\eps)\to\bs X$. Our goal
here is to estimate the parameters that maximize the likelihood
of the observation of $n$ independent samples of $\bs X$ that we
will denote $\bs x^{*n}=(\bs x_1, \ldots, \bs x_n)$.

Let $q_m(\bs z; \theta)$ denote the prior likelihood of the hidden (or missing)
variable~$\bs Z$ that generates $\bss\be$. Denote by $q_c(\bs
x|\bs z;\theta)$ the conditional distribution of~$\bs X$ given\vadjust{\goodbreak}
$\bs Z=\bs z$ which is, in all our models, a Gaussian distribution
centered at the ICA decomposition. The joint density is
\[
q(\bs x,\bs z;\theta) = q_{c}(\bs x|\bs z
; \theta) q_m(\bs z;\theta)
\]
and the marginal distribution of
$\bs X$ has density
\[
q_{\mathrm{obs}}(\bs x;\theta) = \int q_{c}(\bs x|\bs z
; \theta) q_m(\bs z;\theta) \,d\bs z.
\]

Our goal is to maximize the likelihood of the observations, namely, to find
%
%
\begin{equation}
\label{eqMAP}
\hat\theta_n = \mathop{\Argmax}_\theta q_{\mathrm{obs}}^{*n}(\bs
x^{*n}; \theta) \qquad\mbox{with } q^{*n}_{\mathrm{obs}}(\bs x^{*n};\theta
) = \prod_{k=1}^n
q_{\mathrm{obs}}(\bs x_k;\theta).
\end{equation}

\subsection{SAEM algorithm}

\label{secalgo}

This problem can, in principle, be solved using the expectation--maximization (EM) algorithm. With the EM, a local maximum of the
likelihood is computed recursively while replacing the
missing variables with a conditional expectation. For each observation
$\bs
x_k$ and parameter $\theta$, we
define the conditional density of $\bs Z$ by
%
%
\begin{equation}
\label{eqEstep}
\nu_{k, \theta} (\bs z) =
q(\bs z | \bs X = \bs x_k ; \theta) .
\end{equation}

The EM algorithm iterates the following two
steps, where $t$ indexes the current iteration:
\begin{longlist}
\item[$E$: expectation.]
Compute
$\ell_{t+1} \dvtx \theta\mapsto\ell_{t+1}(\theta) = \sum_{k=1}^n
\mathbb
{E}_{\nu_{k,\theta_t}} [ \log
q(\bs x_k, \bs Z ; \theta)].$
\item[$M$: maximization.] Set
$ \theta_{t+1} = \Argmax_{\theta\in\Theta} \ell
_{t+1}(\theta
) $.
\end{longlist}

The models we have discussed for ICA belong to the curved exponential
family, in the sense that the joint
distribution of hidden and observed variables for a given parameter can
be expressed as
\[
\log q(\bs x, \bs z; \theta) = \phi(\theta)\cdot\bs S(\bs x, \bs z) -
\log C(\theta) ,
\]
where $\bs S$ is a multidimensional sufficient statistic, $\phi$ is a
fixed, vector-valued function of the parameters, $C$ is a normalizing
constant and the dot
refers to the usual Euclidean dot product. This implies that
\[
\ell_{t+1}(\theta) = \phi(\theta)\cdot\Biggl( \sum_{k=1}^n
\mathbb{E}_{\nu_{k,\theta_t}} \bs S\Biggr) - n\log C(\theta).
\]
Thus, the $E$-step only requires computing the conditional
expectations of the sufficient statistic, and the $M$-step is
equivalent to maximum likelihood for a fully observed model, with the
empirical expectation of the sufficient statistic equal to
\[
\bar{\bs{S}}_{t+1} = \frac1n \sum_{k=1}^n
\mathbb{E}_{\nu_{k,\theta_t}} \bs S.
\]
This is an important property (satisfied by our models) for the
numerical feasibility of the
EM algorithm.\vadjust{\goodbreak}


However, this is not enough, since one must also
be able to explicitly compute the conditional expectations. For several
of our models, there is no closed form expression for the densities
$\nu_{k, \theta}$. For others, like IFA, for which such an expression
can be derived, its computational complexity is exponential in the
number of components and rapidly becomes intractable (details are
given in the \hyperref[app]{Appendix}).

A common way to overcome this difficulty is to
approximate these conditional distributions by Dirac measures at their
mode. The resulting algorithm is sometimes called EM-MAP or FAM-EM (for ``Fast
Approximation with Mode'') [\citet{AAT}, \citet{aktmfca08}]. At each
iteration of the algorithm, one computes the most likely hidden variables
$\hat{\bs{z}}_k$, $1\leq k\leq n$,
with respect to the current parameters:
%
%
\begin{equation}
\label{eqbeta*}
\hat{\bs{z}}_{t,k} = \mathop{\Argmax}_{\bs z} \bigl[\log\bigl(
q(\bs z|\bs X = \bs x_k, \theta_t)\bigr)\bigr] .
\end{equation}
The $M$-step then maximizes the likelihood for the ``completed
observations''~%
$\bs x^{*n} $ and $\hat{\bs{z}}_{t,1}, \ldots, \hat{\bs{z}}_{t,n}$.

The statistical accuracy of this approximation is unclear,
since it estimates a number of parameters that scales like the number
of observations. Consistency of the obtained estimator when $n$ goes
to infinity cannot be proved in general. Some experimental
evidence of asymptotic bias is demonstrated in Section \ref{secExp} below.

In spite of these remarks, this approach (or approaches similar to it)
is the most common choice for training probabilistic ICA models
[\citet{grNeuralComputation2005}, \citet{hyvarinen},
Olshausen and Field (\citeyear{ofNature1996,of96})]. In the
under-determined problem ($p\gg d$), this algorithm has also been
implemented in \citet{bremondmoulinescardoso}.

Although the conditional distribution is not explicit, it is still
possible (as we shall see later) to sample from it. The conditional
expectation of the sufficient statistics ($\bar{\bs{S}}_{t+1}$) can
therefore be
approximated by Monte Carlo simulation, as proposed in
\citet{TannerBook} and \citet{TannerWei90} with the MCEM
(Monte Carlo EM)
algorithm. The resulting method, however, is heavily
computational. Also, there is no guarantee that the errors resulting
from the approximation to the \mbox{$E$-step} will cancel out to provide an
estimator converging to a local maximum of the likelihood.

In this regard, a more interesting procedure, which has been proposed
in \citet{DLM}, is a stochastic approximation of
the EM algorithm, called SAEM. It replaces the $E$-step by a stochastic
approximation step for the conditional likelihood (or, in practice,
for the conditional expectation of the sufficient statistics), on
which the $M$-step is based. More precisely, based on a sequence $\De_t$
of positive numbers decreasing to 0, the algorithm iterates the
following two steps (assuming the $t$th iteration):
\begin{longlist}
\item[SAE step.]
For $k=1,
\ldots, n$, sample a new hidden variable $\bs z_{t+1,k}$ according to the
conditional distribution $\nu_{k,\theta_t}$ and define
\[
\ell_{t+1}(\theta) = \ell_t(\theta) + \De_t \Biggl(\sum_{k=1}^n
\log
q(\bs x_k, \bs z_{t+1,k} ; \theta) - \ell_t(\theta)\Biggr).
\]
\item[$M$ step.] Set
%
\[
\theta_{t+1} = \mathop{\Argmax}_{\theta\in\Theta} \ell
_{t+1}(\theta).
\]
\end{longlist}
For exponential families, the SAE step is more conveniently (and
equivalently) replaced
by an update of the estimation of the conditional expectation of the
sufficient statistics, namely,
\[
\bar{\bs{S}}_{t+1} = \bar{\bs{S}}_t + \De_t \Biggl(\frac1n \sum
_{k=1}^n \bs S(\bs x_k, \bs
z_{t+1,k}) -\bar{\bs{S}}_t\Biggr)
\]
with
\[
\ell_{t+1}(\theta) = \phi(\theta)\cdot\bar{\bs{S}}_{t+1} - \log
C(\theta)
\]
being maximized in the $M$-step.
Note that this algorithm is
fundamentally distinct from the SEM method [\citet
{CeleuxDiebolt85}] in
which the $E$-step directly defines
$\ell_{t+1}(\theta) = \sum_{k=1}^n \log
q(\bs x_k, \bs z_{t+1,k} ; \theta)$.

A final refinement may be needed in the SAEM algorithm, when directly
sampling from the
posterior distribution is infeasible, or inefficient, but can be done
using Markov Chain Monte Carlo (MCMC) methods. In this situation,
there exists, for each $\theta$ and $\bs x$, a transition probability
$z\mapsto\Pi_{\bs x,\theta}(z, \cdot)$ such that the associated
Markov chain
is ergodic and has the posterior probability $q(\cdot|\bs X
= \bs x;
\theta)$ as stationary distribution. The corresponding variant of the
SAEM (which we shall still
call SAEM) replaces the direct sampling operation
\[
\bs z_{t+1,k} \sim\nu_{k, \theta_t} = q(\cdot|\bs X=\bs x_k, \theta_t)
\]
by a single Markov chain step
\[
\bs z_{t+1,k} \sim\Pi_{\bs x_k, \theta_t}(\bs z_{t,k}, \cdot) .
\]
This procedure
has been introduced and proved convergent for
bounded missing data in \citet{kuhnlavielle}. This result has been
generalized to unbounded hidden random
variables in \citet{aktdefmod}.

To ensure the convergence of this algorithm in the noncompact case
(which is our case in the models above),
one needs, in principle, to introduce a truncation on random boundaries
as in
\citet{aktdefmod}. This would add a new operation
between the stochastic approximation and the maximization steps,
with the following \textit{truncation step}. Let $\mathcal S$ be the range
of the sufficient\vadjust{\goodbreak} statistic, $S$.
Let
$(\Kapa_q)_{q\geq0} $ be an increasing sequence of compact subsets of
$\mathcal{S}$ such as $\bigcup_{q\geq0} \Kapa_q = \mathcal{S} $
and $ \Kapa_q \subset\operatorname{int}(\Kapa_{q+1}) , \forall q
\geq0$.
Let $(\delta_t)_t$ be a decreasing sequence of positive numbers.
If $\bar S_{t+1}$
wanders out of $\Kapa_{t+1}$ or if $| \bar S_{t+1} - \bar S_{t}| \geq
\delta_t$, then the algorithm is reinitialized in a
fixed compact set.

More details can be found in \citet{andrieumoulinespriouret} and
Allassonni{\`e}re, Kuhn and Trouv{\'e} (\citeyear{aktdefmod}). In
practice, however, our algorithms work properly without this technical
hedge.

\subsection{Application to our models}
\label{secmethast}

To complete the description of the SAEM algorithm for a given model, it
remains to make explicit (i) the specific form of the sufficient
statistic $\bs S$; (ii) the corresponding maximum likelihood estimate
for complete observations; and (iii) the transition kernel for the MCMC
simulation. Formulae for (i) and (ii) are provided in the \hyperref
[app]{Appendix} for the ICA models we have described here. For (iii),
we have used a Metropolis--Hastings procedure, looping over the
components (sometimes called ``Metropolis--Hastings within Gibbs
Sampling'') that we now describe. This is for a fixed observation $\bs
x_k$ and parameter~$\theta$, although we do not let them appear in the
notation. So we let $\nu= \nu_{k, \theta}$ be the probability that
needs to be sampled from.

In the Metropolis--Hastings procedure, one must first specify a
candidate transition probability
$\rho(\bs z, \tilde{\bs{z}})$. A Markov chain $(\bs Z_t, t=0, 1,
\ldots)$
can then be defined by the two iteration steps, given $\bs Z_t$:
\begin{longlist}[(1)]
\item[(1)] Sample $\bs z$ from $\rho(\bs Z_t, \cdot)$.
\item[(2)] Compute the ratio
\[
r(\bs Z_t, \bs z) = \frac{\nu(\bs z)\rho(\bs z, \bs Z_t)}{\nu(\bs
Z_t)\rho(\bs Z_t, \bs z)}
\]
and set $\bs Z_{t+1} = \bs z$ with probability $\min(1, r)$
and $\bs Z_{t+1} = \bs Z_t$ otherwise.
\end{longlist}
An interesting special case is when $\rho$ corresponds to a Gibbs sampling
procedure for the prior distribution, $q_m(\mathbf z;\theta)$. Given the
current simulation $\bs z$, one randomly selects one component $z^j$
and generates
$\tilde{\bs{z}}$ by only\vspace*{1pt} changing $z^j$, replacing it by $\tilde
z^j$ sampled from the conditional distribution $q_m(\tilde z^j|z^i,
i\neq j;\theta)$. In this case, it is easy to see that the ratio $r$ is
then given by
\[
r( \tilde{\bs{z}}, \bs z) = \frac{q(\bs{x}_k | \tilde{\bs{z}}
)}{q(\bs{x}_k | \bs{z})}.
\]
The Markov kernel is then built by successively applying the previous
kernel to each component.

Our implementation follows this procedure whenever the current set of
parameters leads to an irreducible transition probability $\rho$. This
is always true, except for the censored models, in which parameters
$\alpha\in\{0,1\}$ or $\ga\in\{0, \frac12\}$ are\vspace*{1pt}
degenerate and must be replaced by some fixed values $\alpha_0$ and~$\ga_0$
in the definition of $\rho$.\vadjust{\goodbreak}

\section{Reconstruction}
\label{secreconstruction}

Assuming that the parameters in the model are known or have been
estimated, the reconstruction problem consists in estimating the
hidden coefficients of the independent components,
$\hat{\bs{\be}}\in\mR^p$, based on a~new observation of $\bs x\in
\mathbb
R^d$. As noticed in the \hyperref[sec1]{Introduction}, this is a separate
problem. Estimating model parameters is based on the likelihood of the
observation, which integrates out hidden variables. In contrast,
reconstructing hidden decomposition vectors from data is typically
done by minimizing a~chosen loss function for a fixed choice of model
parameters, and is based on the posterior likelihood (proportional to
the complete likelihood). Even if this does not constitute our main
focus here, we briefly describe in this section how the MAP estimator,
based on maximizing the complete likelihood, can be achieved using the
models presented in this paper.

Reconstruction with probabilistic ICA models is not as straightforward
as with complete ICA,
for which the operation reduces to solving a linear system. A
natural approach is maximum likelihood, that is, (with our notation) find
$
\hat{\bs{z}} = \arg\max_{\bs z} \phi(\theta)\cdot S(\bs x,
\bs z)
$
and deduce $\hat{\bss{\be}}$ from it.

This maximization is not explicit, although simpler for our first two
models. Indeed, for Log-ICA, this requires minimizing
\[
\frac{1}{2\sig^2} |\bs x - \bs A\bss\be|^2 + 2\sum_{j=1}^p \log
(e^{\be^j} + e^{-\be^j}).
\]
(We take $\bss\mu_0=0$ in this section, replacing, if needed, $\bs x$
by $\bs x - \bss\mu_0$.)

The Laplacian case, Lap-ICA, gives
\[
\frac{1}{2\sig^2} |\bs x - \bs A\bss\be|^2 + \sum_{j=1}^p |\be^j|.
\]

Both cases can be solved efficiently by convex programming. The
Laplacian case is similar (up to the absence of normalization of the
columns of
$\bs A$) to the Lasso regression algorithm [\citet{TibLasso}], and can
be minimized using an incremental procedure on the set of vanishing
$\be^j$'s [\citet{LARS}].

The other models also involve
some form of quadratic integer programming, the general solution of which
being NP-complete. When dealing with large numbers of components, one
must use generally suboptimal optimization strategies (including local
searches) that have
been developed for this context [see \citet{lisun06}, e.g.].

The EG-ICA problem requires minimizing
\[
\frac{1}{2\sig^2} \Biggl|\bs x - \sum_{j=1}^p s^j y^j \bs a_j\Biggr|^2
+ \sum_{j=1}^p s^j + \frac12 \sum_{j=1}^p (y^j-\mu)^2
\]
with $s^1, \ldots, s^p \geq0$. This is not convex, but one can use in
this context an alternate\vadjust{\goodbreak}
minimization procedure, minimizing in $\bs y$ with fixed $\bs s$ and
in $\bs s$ with fixed $\bs y$. The first problem is a straightforward
least squares and the second requires quadratic programming.

The symmetrized IFA model leads to minimize
\[
\frac{1}{2\sig^2} |\bs x - \bs A \bss\beta|^2 + \frac{1}{2} \sum
_{j=1}^p
(\beta^j - b^jm_{t^j})^2 + \sum_{j=1}^p \log w_{t^j}
\]
with respect to $\bss\be$, the unobserved configuration of labels $\bs
t$, and the sign change~$\bs b$. When labels and signs are given, the
problem is quadratic in $\bss\be$. Given $\bss\be$ and $\bs t$, the
optimal $\bs b$ is explicit, and for fixed $\bss\be$ and $\bs b$, the
search for an optimal $\bs t$ reduces to a quadratic integer
programming problem. For small dimensions, it is possible to make an
exhaustive search of all $(2K+1)^p$ possible configurations of labels
and signs.


For the BG-ICA, we must minimize
\[
\frac{1}{2\sig^2} \Biggl|\bs x - \sum_{j=1}^p b^j y^j \bs a_j\Biggr|^2
+ \rho \sum_{j=1}^p b^j + \frac12 \sum_{j=1}^p (y^j-\mu)^2
\]
with $\rho= \log((1-\alpha)/\alpha)$ and $b^j\in\{0,1\}$. The
minimization in $\bs b$ is a $(0,1)$-quadratic programming problem, an
exhaustive search being feasible for small $p$. Given $\bs b$, the
optimal $\bs y$ is provided by
least squares.

Concerning the EBG-ICA, we must minimize
\[
\frac{1}{2\sig^2} \Biggl|\bs x - \sum_{j=1}^p s^j b^j y^j \bs
a_j\Biggr|^2 + \sum_{j=1}^p s^j + \rho\sum_{j=1}^p b^j + \frac12
\sum_{j=1}^p (y^j-\mu)^2
\]
with $\rho= \log((1-\alpha)/\alpha)$, $s^1, \ldots, s^p>0$ and
$b^j\in\{0,1\}$. This is again a $(0,1)$-quadratic programming problem
in $\bs b$ and, given $\bs b$, the optimal $\bs y$ and $\bs s$
are computed similarly to the EG-ICA model.

With ET-ICA, the objective function is
\[
\frac{1}{2\sig^2} \Biggl|\bs x - \sum_{j=1}^p s^j y^j \bs a_j\Biggr|^2
+ \sum_{j=1}^p s^j + \rho\sum_{j=1}^p |y^j|
\]
with $\rho= \log((1-2\ga)/2\ga)$, $y^1, \ldots, y^p\in
\{-1,0,1\}$
and $s^1, \ldots, s^p>0$. This is a quadratic integer
programming in $\bs y$, with a complexity of $3^p$ for an exhaustive
search. Given $\bs y$, computing $\bs s$ is a standard quadratic
programming problem.

The TE-ICA problem, requiring to minimize
\[
\frac{1}{2\sig^2} \Biggl|\bs x - s \sum_{j=1}^p y^j \bs a_j\Biggr|^2 +
s + \rho\sum_{j=1}^p |y^j|
\]
is slightly simpler since, in this case, the computation of $s\geq0$
given $y$ is straightforward.\vadjust{\goodbreak}

The TEoff-ICA model involves a third hidden variable $\mu$.
This leads to the following objective function to minimize both in $s$,
$\bs y$ and $\mu$:
\[
\frac{1}{2\sig^2} \Biggl|\bs x - \bss\mu-s \sum_{j=1}^p y^j \bs a_j\Biggr|^2
+ s + \rho\sum_{j=1}^p |y^j|
\]
with $ \bss\mu= (\mu, \ldots, \mu)\in\R^d$, and $s>0$. Given
$\bss\mu$, the minimization with respect to $s$ and $\bs y$ is done as
in the previous TE-ICA model. The minimization over~$\mu$ has a closed
form:
\[
\mu= \frac{1}{d} \sum_{i=1}^d \Biggl( \bs x^j - s \sum_{j=1}^p y^j \bs
a_{i,j} \Biggr) .
\]

\section{Experiments}
\label{secExp}

\subsection{Synthetic image data}

\subsubsection{Data set}

We first provide an experimental analysis using synthetic data, which
allows us to work in a controlled environment with a~known ground
truth. In this setting, we assume that the true distribution is the
Bernoulli--Gaussian (BG) model, with two components ($p=2$). The
probability $\alpha$ of each component to be ``on'' is set to $0.8$.
We run experiments based on $30, 50$ or $100$ observations, and
vary the standard deviation of the noise using $\sigma=0.1, 0.5,
0.8, 1.5$.

\begin{figure}[b]

\includegraphics{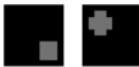}

\caption{Two decomposition images used for sampling synthetic data.}
\label{figToyExampleData}
\end{figure}

\begin{figure}

\includegraphics{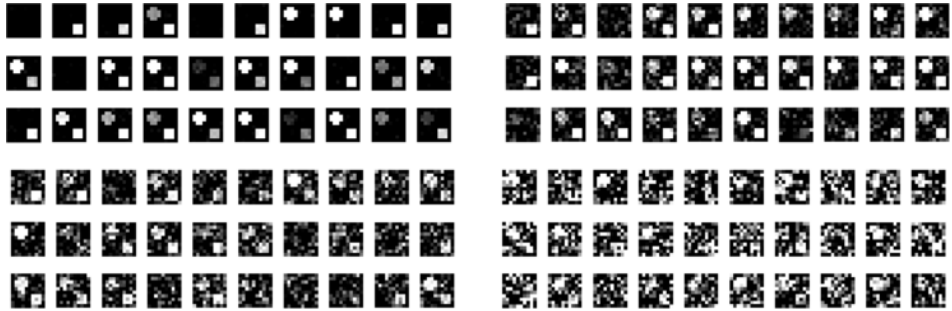}

\caption{Samples of the training sets used for synthetic data with
different level of noise. $\sigma=0.1$, $0.5, 0.8, 1.5$ are upper
left, upper right, lower left and lower right, respectively.}
\label{figToyExampleTrainSet}
\end{figure}

The components are represented as two-dimensional binary images (grey
levels being either $0$ or $1$). The first one is a black image (grey
level equals~$0$) with a white cross (grey level $1$) in the top left
corner. The second one has a~white square (same grey level) in the
bottom right corner. These two images are shown in Figure \ref
{figToyExampleData}. Figure \ref{figToyExampleTrainSet} presents
%
%
$30$ images sampled from this model with the different noise levels.
The training sets were sampled once and used in all the comparative
experiments below.
%
%
We used a fixed color map for all figures to allow for comparisons
across experiments (this explains why the patterns in Figure \ref
{figToyExampleData} appear as grey instead of white).

\subsubsection{Interpretation of the results}

We have compared the following estimation strategies: (1) FAM-EM
algorithm [\citet{grNeuralComputation2005},
\citet{ofNature1996}, \citet{tfNeuralComputation2002}] (which
maximizes the likelihood with respect to parameters and hidden
variables together) with the Log-ICA model (Logistic distribution); (2)\vadjust{\goodbreak}
SAEM with the same Log-ICA model; (3) SAEM for the IFA model, and (4)
EM with the IFA model [\citet{Attias99},
\citet{wwNeuralComputing2001}]; (5) SAEM for the true BG-ICA
model; (6) finally, we also ran a standard ICA decomposition using
fast-ICA [\citet{hoNeuralComputation97}] with a requirement of
computing only two components (with a preliminary dimension reduction
based on PCA). Models (3) and (4) are theoretically equivalent, and our
experiments evaluate how they differ numerically. We reemphasize that
the EM algorithm for the IFA model is only feasible for a reasonably
small number of components,~$p$, and number of mixtures, $K$ (with a
complexity in $K^p$), whereas this limitation does not apply to the
SAEM algorithm (see the \hyperref[app]{Appendix} for more details). For
other alternative approaches to the EM for the IFA model (including the
use of the FAM-EM strategy), see \citet{bwICASSP2005},
\citet{ccoaICMLA2008}, \citet{gsrICCV2003},
\citet{lpIWICABSS2000}, \citet{vsptArXiv09}. The fast-ICA
algorithm used in (6) is nonparametric (and maximizes an approximation
of the negentropy of the model).

We also notice that
(1), which requires minimizing in $A, \sigma^2$ and $\bs b$, is
ill-posed because a transformation $(A, \bs b)\to(\lambda A, \bs
b/\lambda)$ always decreases the likelihood when $\lambda> 1$, which
implies that the optimal $A$ is unbounded. To address this, one
solution is to use a prior distribution for $A$, or enforce some
normalization. We chose the latter option, enforcing the empirical
mean square of all $\bs b$'s to be equal to $\log2$ as implied by the
logistic distribution.

Table \ref{tabmse} provides mean-square errors for the estimation of $A$
based on different models and algorithms, and for different noise
levels and sample size. Each error is computed from 50 repeats of the
full experiment (sampling from the true model followed by
estimation). The mean square error (MSE) is defined by
\[
\mathrm{MSE} = \frac{1}{|\Lambda|}\sum_{x\in\Lambda}
\bigl[\bigl(A_{\mathrm{est}}(x,1)-A_{\mathrm{true}}(x,1)\bigr)^2 +
\bigl(A_{\mathrm{est}}(x,2)-A_{\mathrm{true}}(x,2)\bigr)^2 \bigr],
\]
where $\Lambda$ is the grid of pixels, $|\Lambda|$ its cardinality,
$A_{\mathrm{est}}$ is the estimated decomposition matrix and $A_{\mathrm{true}}$ the
true one (up to a permutation and a change of sign).
The lack of monotonicity in mean squared errors with respect to $\sigma
$ may come from the small number of simulations which are averaged
here. The estimation is to proceed $50$ times but a larger number of
simulations would solve the problem.

%
%
\begin{table}
\tabcolsep=0pt
\caption{Mean-square estimation error for various
combinations of
algorithms and models based on estimations based on $30/100$ samples
and several noise levels. Each mean-square error is an average over
50 independent repeats}\label{tabmse}
\begin{tabular*}{\tablewidth}{@{\extracolsep{\fill}}lcccccccc@{}}
\hline
& \multicolumn{4}{c}{\textbf{30 images per training set}} &
\multicolumn{4}{c@{}}{\textbf{100 images per training set}}\\[-4pt]
& \multicolumn{4}{c}{\hrulefill} &
\multicolumn{4}{c@{}}{\hrulefill}\\
\textbf{Algo/model} & $\bolds{\sigma=0.1}$ & $\bolds{\sigma= 0.5}$
& $\bolds{\sigma= 0.8}$ & $
\bolds{\sigma= 1.5}$ & $\bolds{\sigma=0.1}$ & $\bolds{\sigma= 0.5}$
& $\bolds{ \sigma= 0.8 }$ & $
\bolds{\sigma= 1.5}$\\
\hline
FAM-EM/Log & 0.55 & 0.49 &0.51&0.82 &0.52 & 0.47 &0.46 & 0.62 \\
SAEM/Log & 0.05 & 0.06 & 0.10 & 0.26 & 0.03 & 0.06 &0.06 & 0.11 \\
SAEM/IFA & 0.19 & 0.18 & 0.16 & 0.20 & 0.16 & 0.16 &0.09 & 0.10 \\
EM/IFA & 0.05 & 0.04 & 0.06 & 0.15 & 0.05 & 0.03 &0.03 &0.06 \\
SAEM/BG & 0.09 & 0.13 & 0.16 & 0.6\hphantom{0} & 0.07 & 0.07 & 0.05 & 0.25 \\
\hline
\end{tabular*}
\end{table}

%
%
\begin{table}[b]
\tabcolsep=0pt
\caption{Estimated noise variance with the different models and the
two different algorithms for $30$, $50$ and $100$ images in the
training set. These variances correspond to the estimated
decomposition vectors presented in Figure
\protect\ref{figpicsToyExample30obs}}
%
\label{tabestimatedSigma2}
\begin{tabular*}{\tablewidth}{@{\extracolsep{\fill}}ld{1.4}ccccc@{}}
\hline
& \multicolumn{6}{c@{}}{\textbf{Algo/model}}\\[-4pt]
& \multicolumn{6}{c@{}}{\hrulefill}\\
& \multicolumn{1}{c}{\textbf{True} $\bolds{\sigma^2}$}
& \multicolumn{1}{c}{\textbf{FAM-EM/Log}}
& \multicolumn{1}{c}{\textbf{SAEM/Log}}
& \multicolumn{1}{c}{\textbf{EM/IFA}}
& \multicolumn{1}{c}{\textbf{SAEM/IFA}}
& \multicolumn{1}{c@{}}{\textbf{SAEM/BG}} \\
\hline
30 images in the & 0.001 & 0.0088 & 0.0086 &0.0097 & 0.0089 &
0.0087 \\
\quad training set & 0.2500 & 0.2253 & 0.2224 & 0.2240 & 0.2410 &
0.2226 \\
& 0.6400 & 0.5685 & 0.5577 & 0.5534 & 0.6092 &
0.5569 \\
& 2.2500 & 2.0375 & 1.9978 & 2.1199 &  2.0735 & 2.0009\\
[6pt]
%
50 images in the & 0.001 & 0.0095 & 0.0092 & 0.0095 & 0.0094 &
0.0092 \\
\quad training set & 0.2500 & 0.2400 & 0.2399 & 0.2363 &0.2524 & 0.2399
\\
& 0.6400 & 0.5831 & 0.5798 & 0.6381& 0.6429 &
0.5795 \\
& 2.2500 & 2.1544&2.1377 & 2.2061&  2.2112 & 2.1366\\
[6pt]
100 images in the & 0.001 & 0.0176 & 0.0097& 0.0095 & 0.0098 &
0.0097 \\
\quad training set & 0.2500 & 0.2432 & 0.2459 & 0.2455 & 0.2564 &
0.2456 \\
& 0.6400 & 0.6225 & 0.6282 & 0.6336& 0.6388 &
0.6280 \\
& 2.2500 & 2.1268 & 2.1479& 2.1767 & 2.1970 & 2.1490\\
\hline
\end{tabular*}
\end{table}

We also evaluated the accuracy of the estimation of $\sigma^2$. The
results are presented in Table \ref{tabestimatedSigma2}. A surprising
result is
that $\sigma^2$ is always well estimated even when the decomposition
vectors are not. This is an important observation which indicates that one
should not evaluate the final convergence of any of these algorithms
based on the convergence of $\sigma^2$ only.

%
%
\begin{figure}

\includegraphics{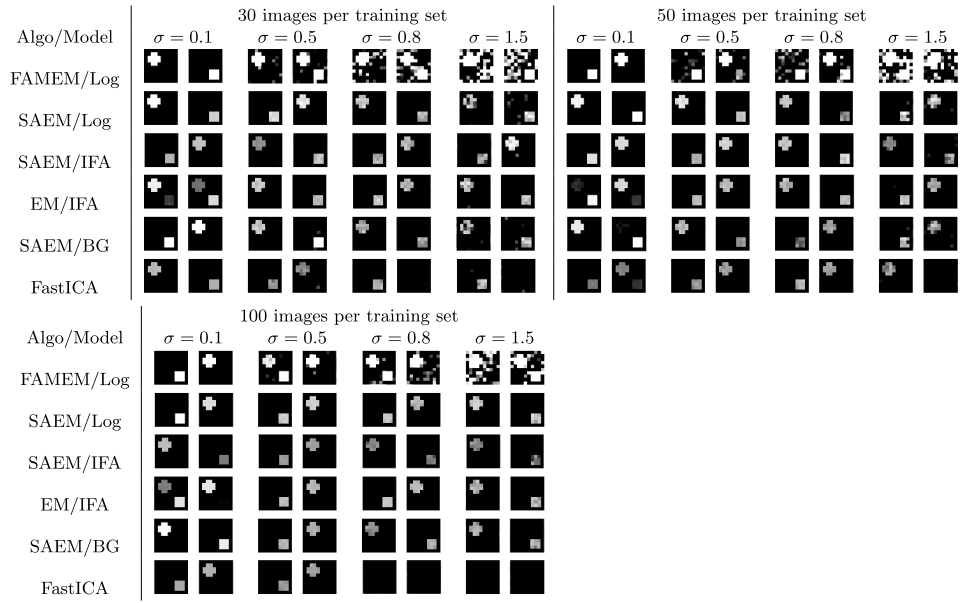}

\caption{Estimated decomposition images with different models
and algorithm. The estimation becomes less and less satisfactory as the
noise variance increases. It is even more pregnant for the FAMEM and
FastICA algorithms for which increasing the number of observation does
not address this problem. The other models estimated using our
algorithm provide similar results and the noise does not drastically
affect the estimation.}
\label{figpicsToyExample30obs}
\end{figure}

A visual illustration of these results is provided in Figure
\ref{figpicsToyExample30obs}, in which a~single (typical) experiment is
displayed for each noise level and sample size.
The algorithms that maximize the likelihood of the observed data
(SAEM, MCEM and EM for the IFA) all provide results that are
consistent with the ground truth,
even when the model used for the estimation differs from the true one.
This statement does not apply to the FAM-EM algorithm
(which maximizes the
likelihood with respect to parameters and hidden variables together),
or to FastICA, which degrade significantly when the noise is high. We
also experienced numerical failures when running the publicly
available software with high noise (we had, in fact, to
resample a new 100-image training set to be able to present results
from this method).

\begin{figure}

\includegraphics{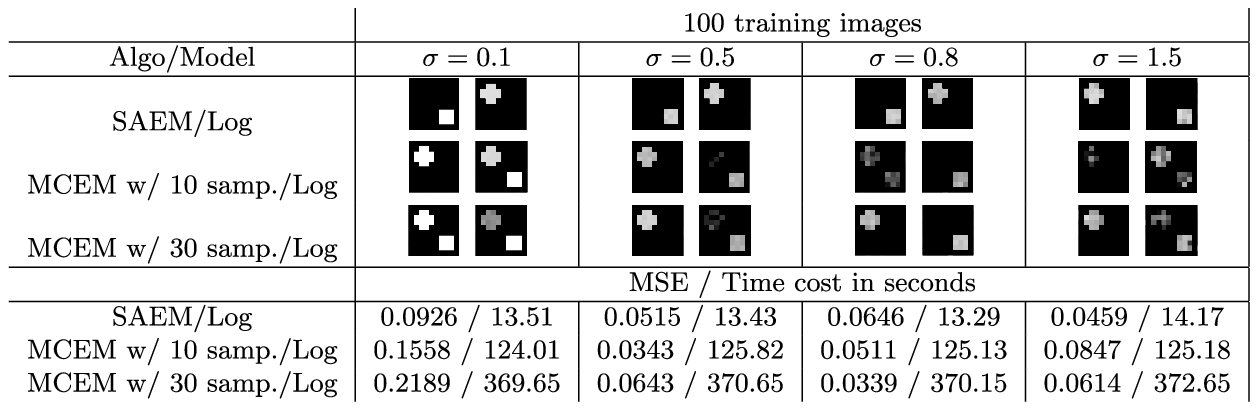}

\caption{Comparison between MCEM and SAEM with the Log-ICA
model. The top images show the decomposition vectors
estimated with either model and for different numbers of
Monte Carlo samples used to approximate the expectation in
the MCEM. The table presents the mean square error (MSE) and
the time cost of each estimation. The results look very
similar (except for low noise variance where the MCEM seems
to behave like the FAM-EM), while the time cost of the MCEM
increases linearly in the sample size.}
\label{figcomparaisonMCEM-SAEM}
\end{figure}

Since these two algorithms both rely on Monte Carlo sampling, we
have compared the performances of our SAEM with a Log-ICA model
and of the Monte Carlo (MC) EM algorithm. The expectation
step of the EM is replaced by an approximation of the expected value
of the sufficient statistics using a Monte Carlo sum. Therefore, at
each iteration of the algorithm, MCEM
requires repeated samples from the posterior distribution of the hidden
variables given the observations. Larger samples yield a better
approximation and generally result in fewer EM iterations to achieve
convergence. Of course, this also implies a computational cost per
iteration which grows linearly in the sample
size. Notice also that we cannot generate independent samples from the
posterior distribution, but only Markov chain samples resulting from
the MCMC sampler described in Section \ref{secmethast}. These
samples are therefore correlated and only asymptotically sample from
the posterior distribution. A comparison of the output of this
algorithm and of the proposed (MCMC-)SAEM is displayed in Figure
\ref{figcomparaisonMCEM-SAEM}. We ran 1,000 iterations, using $10$
and $30$
samples in each Monte Carlo approximation, and the estimation is based
on $100$ observations.
The results are similar, whereas the time cost is about
the number of samples ($10$ or $30$) times longer for the MCEM than
for SAEM. Decreasing the number of samples accelerates the
estimation but degrades the estimations, in particular, when the noise
level is high.

%

We have also made a broad comparison of the required computation time
associated to each algorithm. One must remember, when interpreting
these results, that each algorithm optimizes its own objective
function, and only the ones of the EM/IFA and SAEM/IFA
coincide. While the objective functions of the SAEM models can
be considered as similar, the one associated with the FAM-EM
is quite different, and the comparison must be done with this
in mind.

Another difficulty in computing these numbers is that the true
solution (maximum likelihood, or mode) is unknown, and even if it were
known, all methods are prone to converge to a local maximum and never
get close to it. Because of this, we have used an empirical definition
of the convergence time as the first time at which the maximal
subsequent variation of the current solution is less than $1/1\mbox{,}000$ of
what it was initially. More precisely, if $A(t)$ is the estimated
component matrix at
step $t$, and $d(t)= {\max_{t'\geq t}} |A(t') - A(t_{\max})|$, the
convergence time defined as
%
%
\begin{equation}
\label{eqtconv}
t_{\mathrm{conv}} = \min\{t\dvtx d(t) \leq d(1)/1\mbox{,}000\}\vspace*{-2pt}
\end{equation}
($t_{\max}$ being the maximal number of iterations, equal to
5,000 in our experiments).

These results are summarized in Table \ref{tabcost}. As expected, the
times per iteration of the SAEM-based methods are much smaller than
%
%
\begin{table}
\caption{Comparison of computation costs. Second
column: average
time, in seconds, for 1,000 iterations of each algorithm. Columns 3
to 6: average number of iterations to achieve convergence}\label{tabcost}
\vspace*{-3pt}
\begin{tabular*}{\tablewidth}{@{\extracolsep{\fill}}lcd{6.0}d{6.0}d{6.0}d{6.0}@{}}
\hline
& & \multicolumn{4}{c@{}}{\textbf{Number of iterations to convergence}}\\[-4pt]
& & \multicolumn{4}{c@{}}{\hspace*{-1.5pt}\hrulefill}\\
\textbf{Algo/model} & \multicolumn{1}{c}{\textbf{Time for 1,000 iterations}}
& \multicolumn{1}{c}{\hspace*{-3pt}$\bolds{\sigma=0.1}$} & \multicolumn{1}{c}{$\bolds{\sigma= 0.5}$}
& \multicolumn{1}{c}{$\bolds{\sigma= 0.8}$} & \multicolumn{1}{c@{}}{$\bolds{\sigma=1.5}$}\\
\hline
FAM-EM/Log & \hphantom{0,}206 s & 3\mbox{,}700 & 3\mbox{,}000 & 1\mbox{,}300 & 600 \\
MCEM/Log & \hphantom{0,}140 s & (4\mbox{,}800) & (4\mbox{,}900) & (5\mbox{,}000) & (5\mbox{,}000) \\
SAEM/Log& \hphantom{00,}14 s & 230 & 980 & 1\mbox{,}720 & 2\mbox{,}390 \\
EM/IFA & 1,600 s & 4\mbox{,}400 & 350 & 140 & 50 \\
SAEM/IFA & \hphantom{00,}33 s & 1\mbox{,}510 & 2\mbox{,}240 & 2\mbox{,}920
& 3\mbox{,}670 \\
SAEM/BG & \hphantom{00,}26 s & 530 & 1\mbox{,}210 & 1\mbox{,}900 & 3\mbox{,}660 \\
\hline
\end{tabular*}
\vspace*{-3pt}
\end{table}
with other approaches. This is only partially compensated by an
increased number of
iterations in order to achieve convergence. Note that, in this table,
the number of steps to convergence for the MCEM is close to
$t_{\max} = 5\mbox{,}000$, which indicates that (\ref{eqtconv})
has not been satisfied before the maximal number of
iterations. Note also that the studied model, with two independent
components, is the most favorable for the EM/IFA algorithm, which
would become intractable with a higher number of components.

Another interesting (and difficult to explain) observation from this
table is that the
deterministic algorithms (FAM-EM and EM for the IFA) seem to require
fewer iterations at high noise level, while the trend is opposite for
the stochastic methods. A final remark is that the fastICA algorithm
is much faster than any of these methods when run after reducing the
model dimension using PCA.\vspace*{-3pt}

\subsection{Effect of the number of estimated components}

\label{secnbcomp}
\begin{figure}

\includegraphics{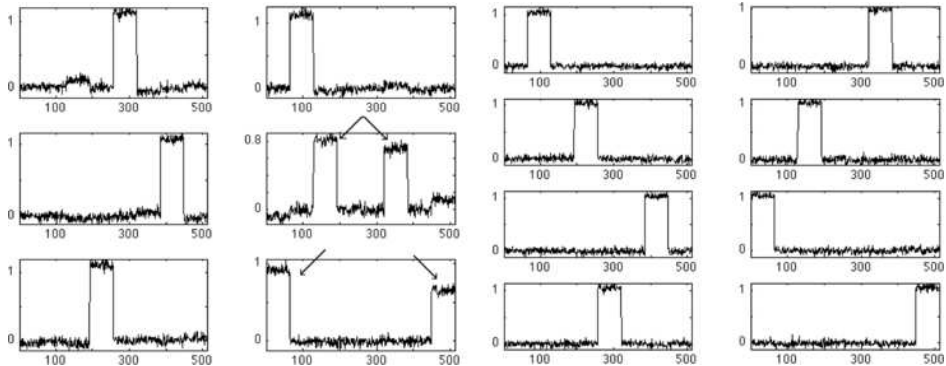}%
\vspace*{-3pt}
\caption{Estimated components with probabilistic ICA. The sample
contains 1,000 signals generated by a shifted Bernoulli--Gaussian model
(see Section \protect\ref{secmean}) with 8 components (the
components being
represented as indicators of 8 nonoverlapping intervals). The true
value of $\alpha$ is $0.5$, and we took $\mu=2$. Left: components estimated
with $p=6$. Right: components estimated
with $p=8$. When estimating only $6$ components, two sources appear in
the same component which will make them always appear together with the
same weight. This is what can be seen pointed by the arrows. However,
when estimating $8$ components, the $8$ sources are recovered.}
\label{figEstimComp6et8}\vspace*{-3pt}
\end{figure}

\begin{figure}

\includegraphics{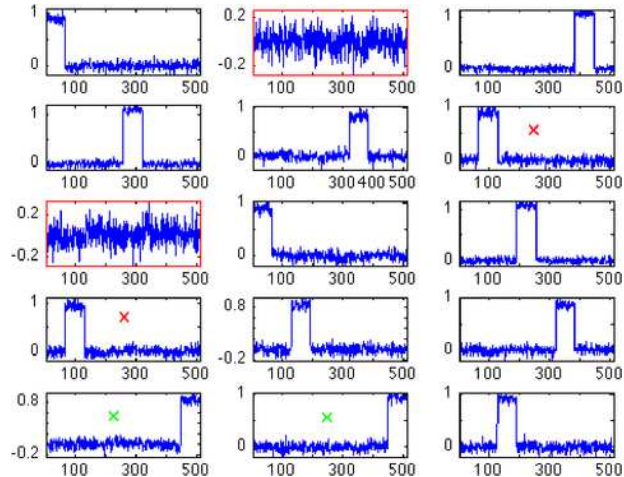}%
\vspace*{-3pt}
\caption{Estimated components with probabilistic ICA. The sample
contains 1,000 signals generated by a shifted Bernoulli--Gaussian model
(see Section \protect\ref{secmean}) with 8 components (the components
being represented as indicators of 8 nonoverlapping intervals). The
true value of $\alpha$ is $0.5$, and we took $\mu=2$. Components
estimated with $p= 15$. We can see that $15$ is too many since among
the $15$~sources found, we can recognize some noise (squared by red
rectangles) and repeated components (red crosses and green crosses are
similar with each other).} \label{figEstimComp}\vspace*{-3pt}
\end{figure}

\begin{figure}[b]

\includegraphics{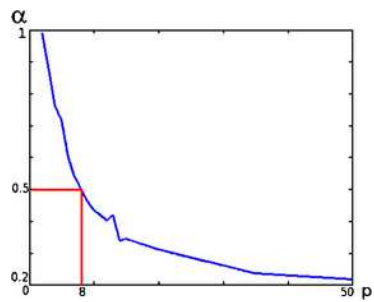}%
\vspace*{-3pt}
\caption{Estimated component activation probability ($\alpha$)
as a function of the model size for a Bernoulli Gaussian
model estimated on the 1,000 signals of a shifted
Bernoulli--Gaussian model. Ground truth is $p=8$ and $\alpha= 0.5$ (red
point).}
\label{figEvolutionGamma}\vspace*{-3pt}
\end{figure}

We now illustrate, with a different model, how, for censored models,
the estimation of the
censoring coefficient evolves with the number of components. In this
experiment we have generated 1,000 samples of a shifted
Bernoulli--Gaussian model (see Section \ref{secmean}) with 8
components (the components being
represented as indicators of 8 nonoverlapping intervals). The true
value of $\alpha$ is $0.5$, and we took $\mu=2$. In Figure
\ref{figEvolutionGamma} we plot the value of the estimated $\alpha$
as a~function of the number of components in the model, $p$. We can see
that this value seems to\vadjust{\goodbreak} decrease to zero, at a rate which is,
however, not linear in $1/p$. The expected number of nonzero
components grows from $2$ for $p=2$, to $8$ when
$p=8$ (correct value---pointed in red in the plot), to about 10 when $p=50$.
The estimated components for $p=6$, $8$ and $15$ are plotted in Figures
\ref{figEstimComp6et8} and \ref{figEstimComp}. This illustrates the
effect of under-dimensioning
%
%
the model, in which some of the estimated components must share some
of the features of several true components (pointed by arrows), and of
over-dimensioning,
in which some of the estimates components are essentially noise
(clearly indicating overfitting of the data---in red rectangles),
while some other
estimated components, which correspond to true ones, are essentially
repeated (twice for the components marked with red and green crosses).
Components are correctly estimated when the estimated model coincides
with the
true model ($p=8$).

%

Although we are not addressing the estimation of the number of
components in
this paper, these results clearly indicate that this issue is
important. We refer the reader to standard approaches in this context,
based on penalizing model complexity, using, for example, the Akaike
Information Criterion (AIC) [\citet{akaike2003}] or the Bayesian
Information Criterion (BIC) [\citet{mcmm2009},
\citet{schwarz78}]. Using a Bayes
prior on parameters would be possible, too, with a straightforward
adaptation of the SAEM algorithm.\vspace*{-3pt}

\subsection{Handwritten digits}

\begin{figure}

\includegraphics{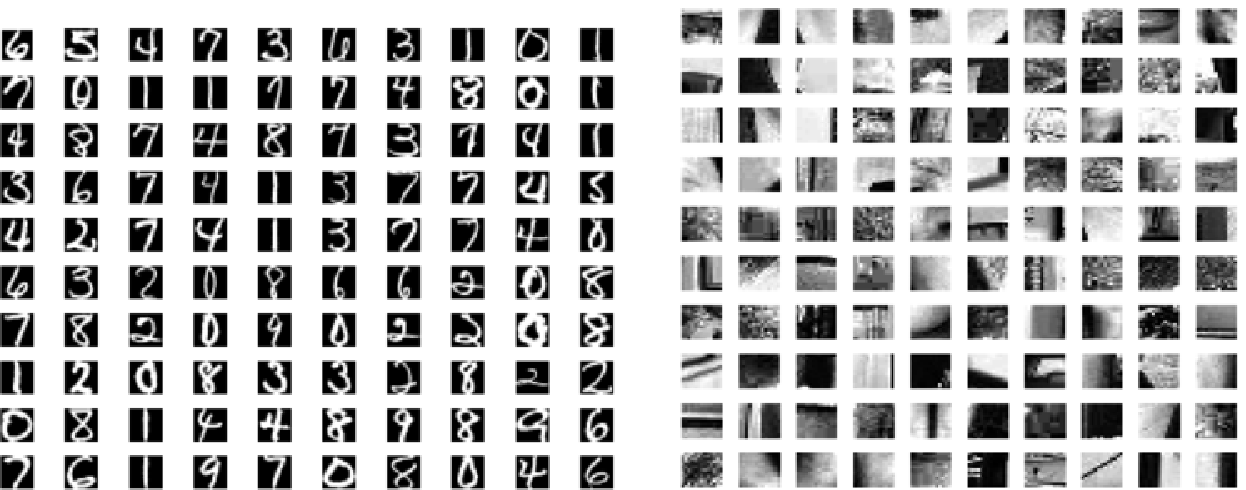}

\caption{100 images randomly extracted from the USPS
database (left) and from the face category in the \textit{Caltech101} data
set (right).}
\label{figtrainUSPS}
\end{figure}

We now test our algorithms on some 2D real images. The first training set
we use is the USPS database, which contains 7,291\vadjust{\goodbreak}
grey-level images of size $16 \times16$. We used the whole database as
a~training set and
computed $20$ decomposition vectors. Some images from this data set are
presented
in Figure~\ref{figtrainUSPS}(left).

%

%
\begin{figure}[b]

\includegraphics{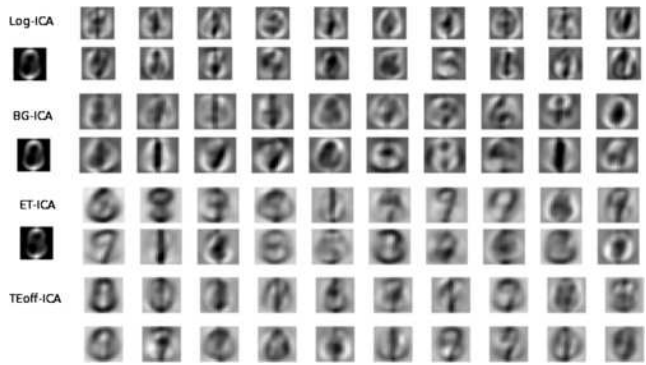}

\caption{Results of the independent component estimation on the USPS
database using four selected models. The training set is composed of
7,291 images containing the 10 digits randomly spread. Left column:
mean image $\mu_0$. Right column: 20 estimated decomposition vectors.
(See Figure 1 in the supplementary file [Allassonni{\`e}re and
Younes (\protect\citeyear{SupfileAY2011})]
for a~larger image.)}
\label{figICAUSPS}
\end{figure}

The different decomposition vectors and the estimated means (when it is
a parameter) are presented in Figure
\ref{figICAUSPS}. Each $10$ by $2$ set of $20$ images on the right column
corresponds to one run of the algorithm for a given model (selecting
the most representative results).

%

Interestingly, the results highlight the advantage of the censored
models compared to the continuous\vadjust{\goodbreak} ones in such situations.
Modeling component coefficients that can vanish with positive
probability (such in BG and ET-ICA) enables to have decompositions
which do not involve vectors shared by all the training sample.
Considering a data set such as USPS, one image in one class is not
easily expressed as a mixture of images from other classes. Therefore,
it is not appropriate to express it as a linear combination of all the
decomposition vectors with nonzero coefficients. This means that we
expect the decomposition vectors to be separated digits and appearing
only for samples belonging to the corresponding class.
This is what we see with the censored models (Figure \ref
{figICAUSPS}, lines 2 to 4), many decomposition vectors represent
well-formed digits, whereas decomposition vectors for other models
(Figure \ref{figICAUSPS}, line 1) mix several digits more often to be
able to cancel the nonexpected features.
These binary or ternary models seem to be very adequate in such situations.

Note that the USPS data set does not have the same number of images of each
digit. There are about twice as many $0$'s or $1$'s as other
digits. This fact explains the ``bias'' one can see on the
mean, on which the shape of the zero is noticeable. In all experiments,
the trace of each digit can be (more or less easily) detected in at
least one of the components, at the exception of
digit~$2$. This is probably due to the large
geometrical variability of the~$2$'s, which is much higher than other digits
(changes of topology-loop or not, changes in global shape) and
therefore difficult to capture.

\subsection{Face images}

\begin{figure}[b]

\includegraphics{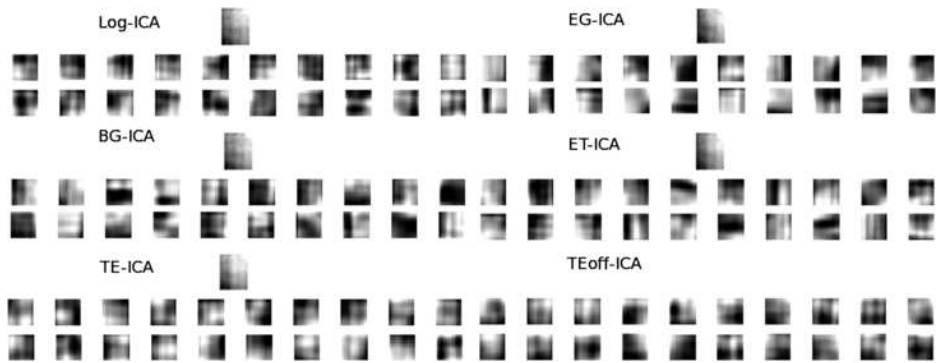}

\caption{Decomposition vectors from six selected models. From left to
right and top to bottom: Log-ICA, Lap-ICA, EG-ICA, BG-ICA, EBG-ICA, ET-ICA,
TE-ICA, TEoff-ICA. For each model the top row is the mean image and the
bottom rows are the 20 corresponding decomposition vectors. (See Figure
2 in the supplementary file [Allassonni{\`e}re and
Younes (\protect\citeyear{SupfileAY2011})] for a larger image.)}
\label{figICFaces101}
\end{figure}

We have run a similar experiment on a data set of face images
(extracted from the Caltech101 data set). Each of these images has been
decomposed into patches of size $13\times13$, with some of them
presented in Figure~\ref{figtrainUSPS}(right). The resulting database
contains 499,697 small images and we estimated $20$ decomposition
vectors. Results are presented in Figure~\ref{figICFaces101}. The
%
%
patterns which emerge from the estimations are quite similar from one
model to another: vertical, horizontal and diagonal separation of the
image into black and white, blobs, regular texture like a regular mesh,
etc.

%
\begin{figure}

\includegraphics{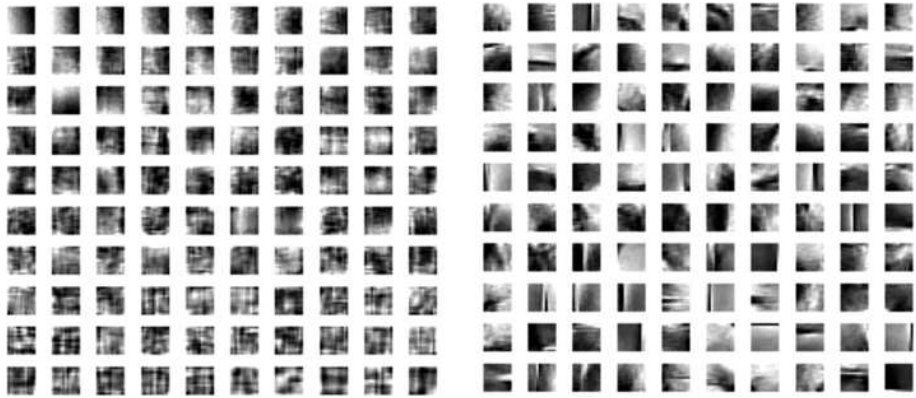}

\caption{$100$ decomposition vectors from 2 models. Left: Log-ICA.
Right: BG-ICA. (See Figure 3 in the supplementary file
[Allassonni{\`e}re and
Younes (\protect\citeyear{SupfileAY2011})] for a larger image.)}
\label{figICFaces101100comp}
\end{figure}

We also ran the same estimation with two of the previous models
looking for $100$ decomposition vectors. The results are presented in
Figure \ref{figICFaces101100comp}. We selected the Log and BG-ICA
since one has a continuous density and the second has a semi-discrete
one. The results are rather different. While the Log-ICA model tends
to capture some textures, the BG-ICA captures some shapes.
In this example, as well as with the digit case, the sparsity of the
decomposition makes sense and plays an important role. This database
is composed of discrete features which can hardly be approximated by
a~linear combination of continuous patterns. The models generating sparse
representations again seems to be better adapted to this kind of data.

\subsection{Anatomical surfaces}

We finally consider a data set containing a~family of 101 hippocampus
surfaces that have been registered to a fixed template using Large
Deformation Diffeomorphic Metric Mapping [Miller, Trouve and Younes
(\citeyear{mty02,eulerlagrange2}), \citet{tro98},
\citet{ty02}]. We here analyze the logarithm of the Jacobian
determinant of the estimated deformations, represented (for each image)
as a scalar field over the surface of the template, described by a
triangulated mesh. These vectors have fixed length ($d=3\mbox{,}223$) equal to
the number of vertices in the triangulation.

The 101 subjects in the data set are separated in 3 groups
with $57$, $32$ and $12$ patients, containing
healthy patients in the first group and patients with Alzheimer's
disease and semantic dementia (denoted the AD group later) at different
stages in the last two groups.

\begin{figure}[b]

\includegraphics{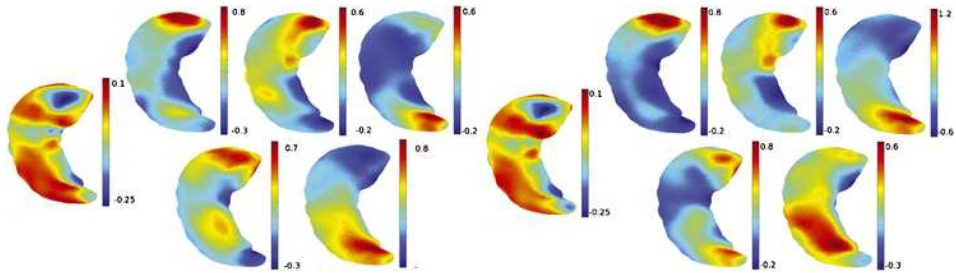}

\caption{Left: mean (left) and 5 decomposition vectors estimated
with the Log-ICA model. Right: mean (left) and 5
decomposition vectors estimated with the Lap-ICA model. Each image
has its own color map to highlight the major patterns.}
\label{figL-ICAhippo}
\end{figure}

\begin{figure}

\includegraphics{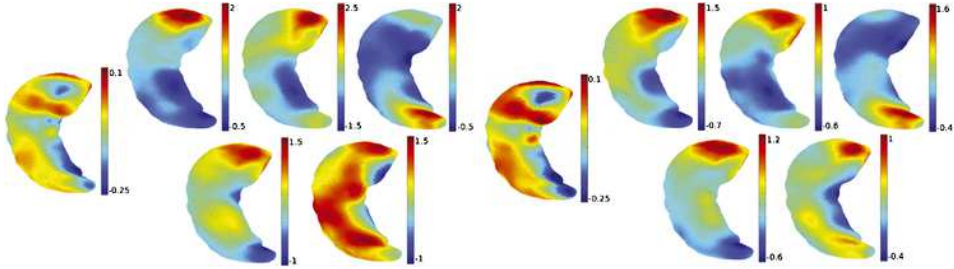}

\caption{Left: mean (left) and 5 decomposition vectors estimated
with the EG-ICA model. Right: mean (left) and 5
decomposition vectors estimated with the BG-ICA model. Each image
has its own color map to highlight the major patterns.}
\label{figEGp-ICAhippo}
\end{figure}

\begin{figure}[b]

\includegraphics{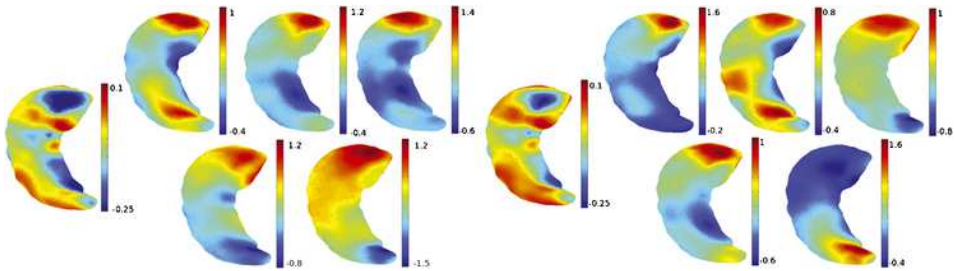}

\caption{Left: mean (left) and 5 decomposition vectors estimated
with the ET-ICA model. Right: mean (left) and 5
decomposition vectors estimated with the TE-ICA model. Each image
has its own color map to highlight the major patterns.}
\label{figTE-ICAhippo}
\end{figure}

Using our algorithm, we have computed $p=5$ decomposition vectors
based on the complete data set. Figures \ref{figL-ICAhippo} to
\ref{figTE-ICAhippo} present these decomposition vectors mapped on
the meshed hippocampus for six selected models. The estimated mean
is shown on the left side and the five corresponding decomposition
vectors are on the right-hand side.
Images are presented with different color maps to facilitate the
visualization of the patterns. In
particular, even if the means seem to contain a lot of information,
their intensities
vary on a very small scale compared to all the decomposition vectors
(they are actually close to $0$).

Although results vary with the chosen model, we can see common
features emerging. First of all, the means are very similar to each
other. The patterns which we can notice on each of them is the
same. For example, there is a~noticeable contraction on the top part
and an extension on the bottom left-hand side of the shape. These
deformations, however, have a small amplitude and can be interpreted as
the ``bias'' of the training set with respect to the template. Concerning
the decomposition vectors
themselves, the pattern of the first vector of the Logistic model is
present in all other models [e.g., in position 1 for the
Laplacian, EG, TE and TEoff models (not shown here), 4 for the BG
model, 5 for the EBG (not shown here)
and~2 for the ET model]. Other patterns occur also, like a contraction
or a growth of the tail part [in vector 3 of Log, Lap, EG, BG, EBG,
TEoff (not shown here) and~5 of TE] or on the bottom of the left part
of the image [in
vectors 4 and 5 of Log, 5 of Lap, EG, BG and TEoff (not shown here) and
in vector 1
otherwise]. These common features seem to be characteristic
of this population.

%
%
\begin{table}
\caption{Mean and standard deviation of the $p$-values for the eight
models with the five decomposition vectors shown in Figures \protect\ref
{figL-ICAhippo} to
\protect\ref{figTE-ICAhippo}. The mean and the standard deviation are
computed over
50 samples of the posterior distributions of the hidden variables to
separate the first group (Control)
with respect to the two others (AZ)}
\label{tabmeanPval}
\begin{tabular*}{\tablewidth}{@{\extracolsep{\fill}}lccccc@{}}
\hline
\textbf{Model} & \multicolumn{1}{c}{\textbf{Log-ICA}}
& \multicolumn{1}{c}{\textbf{Lap-ICA}} & \multicolumn{1}{c}{\textbf{EG-ICA}}
& \multicolumn{1}{c}{\textbf{BG-ICA}} & \multicolumn{1}{c@{}}{\textbf{EBG-ICA}} \\
\hline
Mean on log $ 10^{-3} \times$ & $0.31$ &$ 0.29$ & $0.27$ & $0.33$ &
$0.9$ \\
Std deviation on log $10^{-3} \times$ & $0.16$ & $0.19$ & $0.12$ & $
0.25$ & $1.2$ \\
\end{tabular*}
\begin{tabular*}{\tablewidth}{@{\extracolsep{\fill}}lccc@{}}
\hline
\textbf{Model} & \multicolumn{1}{c}{\textbf{ET-ICA}}
& \multicolumn{1}{c}{\textbf{TE-ICA}} & \multicolumn{1}{c}{\textbf{TEoff-ICA}} \\
\hline
Mean on log $10^{-3} \times$ & $0.27$ & $2.4$ & \hphantom{0}$75.7$ \\
Std deviation on log $10^{-3} \times$ & $0.14$ & $ 2.9$ & $126.2$ \\
\hline
\end{tabular*}
\end{table}

Even if a~careful justification of the following statements would
require a~more thorough study, which would fall out of the scope of the
present paper, these ICA patterns seem to correlate with anatomical
hippocampus regions,
such as those introduced in \citet{hippodata} and \citet
{hipposegment},
in the sense that the supports of the decomposition vectors are located
within subregions of the anatomical segmentation. For example, the
first and
third components from the log-ICA decomposition significantly overlap
with what authors in \citet{hipposegment} refer to as the hippocampus
\textit{lateral zone}, while components 3 and 5 are contained in the
\textit{superior zone}, and component 2 in the \textit{interior-medial
zone}. Similar conclusions apply with most decomposition vectors
obtained with other ICA methods.

In Tables \ref{tabmeanPval} and \ref{tabmeanPval2} we provide the
$p$-values obtained from the comparison of the five ICA coefficients
($\beta$) among the three subgroups. The test is based
on a Hoteling $T$-statistic evaluated on the coefficients, the
$p$-value being computed using permutation sampling.
The test is performed for two different comparisons: first we compare
the healthy group with respect to the two pathological groups. This is
what is shown in Table \ref{tabmeanPval}. The second test compares
the healthy group with the group of 32 mild AD patients. The results
are presented in
Table \ref{tabmeanPval2}.

%
\begin{table}
\caption{Mean and standard deviation of the $p$-values for the eight
models with the five decomposition vectors shown in Figures \protect\ref
{figL-ICAhippo} to
\protect\ref{figTE-ICAhippo}. The mean and the standard deviation are
computed over
50 samples of the posterior distributions of the hidden variables to
separate the first group (Control)
with respect to the second one (mild AZ)}
\label{tabmeanPval2}
\begin{tabular*}{\tablewidth}{@{\extracolsep{\fill}}lccccc@{}}
\hline
\textbf{Model} & \multicolumn{1}{c}{\textbf{Log-ICA}}
& \multicolumn{1}{c}{\textbf{Lap-ICA}} & \multicolumn{1}{c}{\textbf{EG-ICA}}
& \multicolumn{1}{c}{\textbf{BG-ICA}} & \multicolumn{1}{c@{}}{\textbf{EBG-ICA}} \\
\hline
Mean on log $10^{-3} \times$ & $9.0$ & $9.6$ & $8.3$ & $10.9$ & $18.7$
\\
Std deviation on log $10^{-3} \times$ & $3.8$ & $4.8 $ & $2.7$ & \hphantom{0}$
7.6$ & $17.7$ \\
\end{tabular*}
\begin{tabular*}{\tablewidth}{@{\extracolsep{\fill}}lccc@{}}
\hline
\textbf{Model} & \multicolumn{1}{c}{\textbf{ET-ICA}}
& \multicolumn{1}{c}{\textbf{TE-ICA}} & \multicolumn{1}{c@{}}{\textbf{TEoff-ICA}} \\
\hline
Mean on log $10^{-3} \times$& $8.9$ & $30.8$ & $148.7$ \\
Std deviation on log$10^{-3} \times$ & $4.6$ & $28.8$ & $160.4$ \\
\hline
\end{tabular*}
\end{table}

Because SAEM is stochastic and only expected
to converge to a critical point of the likelihood (which may not be
unique), different runs of the algorithm starting from the same initial
point can lead to different limits.
To evaluate the effect of this variability, we ran the algorithm
for each model $50$ times, with the same initial conditions, and
computed an
average and a~standard deviation of the $p$-values.

The results are mostly significant. Indeed, almost all methods yield
$p$-values under $1\%$ when we compare the control population to the AD
groups and less than $3\%$ for the comparison of the control versus
mild AD.

The only model which does not yield significant $p$-values is the
offset case. Both the mean and standard deviation are high (even
higher when we focus on the mild AD population). This suggests that
this model on this database is unstable. One run can lead to
significant decomposition vectors
and a~second one can lead to very different
results. This particular model, which worked well with the USPS
database, for example, does not seem to be adapted to this
type of data that is considered here. The
mean is very close to zero and is therefore not a relevant variable
for this application. The additional variability in the model may have
an adverse effect on the estimation.
In cases where the dimension of the data is much larger than the number
of samples in the training set, it is natural to think that adding more
variability in the estimation process may lead to unstable results and
therefore large variance of estimated parameters. Depending on this
paradigm, the user may prefer to reduce the number of random variables
to the decomposition vector weights only.

%
%

%
%
\begin{figure}

\includegraphics{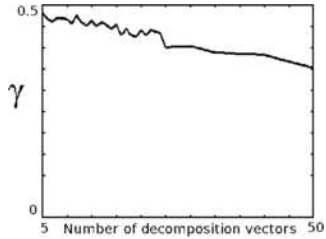}

\caption{Evolution of the probability of one component to activate or
inhibit the corresponding decomposition vector in the ET-model with
respect to the number of decomposition vectors. The training set is the
set of $101$ hippocampi.}
\label{figEvolutionGammaHippo}
\end{figure}

Figure \ref{figEvolutionGammaHippo} provides some insight in the way
components are turned on/off
by the ET-ICA model,\vspace*{-1pt} by plotting the estimated probability,\vadjust{\goodbreak} $\gamma=
P(Y^j_k=-1) = P(Y^j_k=1)$, against the number of decomposition
vectors, $p$. As already noticed in Section~\ref{secnbcomp}, for
small $p$, all components are needed, yielding
$\gamma\simeq1/2$. When more components are added, they do not need
to appear all the time, yielding a decreasing value of $\gamma$.

%
%

%

\section{Conclusion and discussion}

This paper presents a new solution for probabilistic independent
component analysis. Probabilistic ICA enables to estimate a small
number of
features (compared to the dimension of the data) which characterize
a data set. Compared to plain ICA, this avoids the instability of the
computation of the
decomposition matrix when the number of observations is much smaller
than their dimension. We have
demonstrated that the
stochastic approximation EM algorithm is an efficient and
powerful tool which provides a convergent method that estimates the
decomposition matrix. We have shown that this procedure does not
restrict the large choice of distributions for the independent
components, as illustrated by eight models with different
properties, mixing continuous and discrete probability
measures, that we have introduced and studied.

Future works will be devoted to the analysis of nonlinear generative
models that allow for the analysis of data on Riemannian
manifolds, including the important case of shape spaces in which the
models generate nonlinear deformation of given
templates. Generalizations of the methods proposed in
\citet{AAT} and \citet{aksaemmulti} will be developed, in
order to
estimate both
the templates and the generative parameters.

%
%

%

\begin{appendix}\label{app}
\section{Proof of the subexponential tail of the EG-distribution}

Let $(Y,S)$ be a pair of independent random variables where $Y$ and $S$
have a standard normal distribution and an exponential distribution,
respectively.
Let $\beta=Y S$ and assume\vadjust{\goodbreak} $t>0$ so that $\be> t$ implies $Y>0$. We
have [letting $C= (2\pi)^{-1/2}$]
\begin{eqnarray*}
\mathbb P (\beta>t) &=& \mathbb P (s>t/y, y>0)
= C\int_0^\infty\mathbb P \biggl(s>\frac{t}{y}\biggr) \exp
\biggl(-\frac12{y^2}\biggr)\, dy\\
&=& C \int_0^{\infty} \exp\biggl( -\frac12 y^2 -\frac{t}{y}
\biggr)\, dy .
\end{eqnarray*}\vfill\eject
Let $h_t(y) = \frac12 y^2 +\frac{t}{y}$. We can write, letting $z =
y/t^{1/3}$,
\[
h_t(y) = \frac32 t^{2/3} + t^{1/3} \alpha(z)
\]
with $\alpha(z) = ((z-1)^2 + z^{-1}+z-2) /2$. Making the change of
variables $y\to z$ in the integral yields
\[
\mathbb P (\beta>t) = C t^{1/3}e^{-(3/2) t^{2/3}} \int_0^\infty
e^{-t^{1/3} \alpha(z)} \,dz.
\]
Using Laplace's method, we find that the second integral is equivalent
to $\sqrt{2\pi/(6t^{1/3})}$, proving that $\mathbb P(\beta>t$ decays
like $t^{1/6} \exp( -3 t^{2/3}/2)$ when $t\to+\infty$. Note that the
density of $\be$, which is $ g(\be) = \int_0^\infty\exp( -\frac12 y^2
-\frac{\be}{y} ) \,\frac{dy}y $, has a singularity at $\be=
0$.

\section{Maximum likelihood for the complete models}
The $M$-step in our models requires solving the equation
$
E_\theta(\bs S) = [\bs S]
$
where~$[\bs S]$ is a prescribed value of the sufficient statistic (an
empirical average for complete observations, or what we have denoted
$\bar{\bs{S}}_t$ in the $M$-step of the learning algorithm). In the next
sections we provide the expressions of $\bs S$ for the family of models we
consider and give the corresponding solution of the maximum likelihood
equations. Notice that these are closed-form expressions, ensuring the
simplicity of each iteration of the SAEM algorithm.
\subsection{Log-ICA and Lap-ICA models}

$\!\!\!$For these models, the log-likelihood~is
\[
-\sum_{j=1}^p \xi(\be^j) - \frac{1}{2\sigma^2} \Biggl|X -\bss\mu_0-
\sum_{j=1}^p \be^j \bs a_j\Biggr|^2 - \log C(\sig^2, \bs A),
\]
where $\xi(\be) = 2\log(e^\be+ e^{-\be})$ in the logistic case, and
$\xi(\be) = |\be|$ in the Laplacian case. As customary, and to lighten
the formulae, we let $\be^0 = 1$ and $\bs a_0 = \bss\mu_0$, so that
$\bss\be$ and $\bs A$ have size $d+1$, and remove $\bss\mu_0$ from the
expressions for this model and the following ones. We will also leave
to the reader the easy modifications of the algorithms in the case of
shifted models described in Section \ref{secmean}.

The likelihood
can be put in exponential form using the sufficient statistic $\bs S
= (\bss\beta\bss\beta^T, \bs X\bss\beta^T)$, from which the maximum
likelihood
estimator can be deduced using
\[
\cases{
\bs A = [\bs X\bss\be^T]([\bss\be\bss\be^T])^{-1} ,\cr
\sigma^2 = \dfrac{1}{d} (
[|\bs X|^2] -2 \langle\bs A,[\bs X\bss\be^T]\rangle_F + \langle\bs
A^T\bs A,[\bss\be\bss\be^T]\rangle_F = [|\bs X - \bs A\bss\be|^2]/d
) ,}
\]
where $\langle\cdot,\cdot\rangle_F$ refers to the Frobenius dot
product between matrices (the sum of products of coefficients).

\subsection{EG-ICA model}
The likelihood is
\[
-\frac12 \sum_{j=1}^p (Y^j)^2 - \frac{1}2 \sum_{j=1}^p s^j -
\frac{1}{2\sigma^2} \Biggl|X - \sum_{j=1}^p s^j Y^j \bs a_j\Biggr|^2 -
\log C(\sig^2, \bs A)
\]
with sufficient statistic $\bs S = (\bss\be\bss\be^T, \bs X
\bss{\beta}^T)$ with $\be^j = s^j Y^j$.
The maximum likelihood then is
\[
\cases{
\bs A = [\bs X\bss\be^T]([\bss\be\bss\be^T])^{-1},\vspace*{2pt}\cr
\sigma^2 = [|\bs X - \bs A\bss\be|^2]/d.}
\]

\subsection{IFA model}
The complete $\log$-likelihood of the Independent Factor Analysis
model for a single observation $X$ is
\[
-\frac{1}{2\sigma^2} \Biggl|X - \sum_{j=1}^p \be^j \bs
a_j\Biggr|^2 - \frac12 \sum_{j=1}^p (\be^j - b^j m_{t^j})^2 +
\sum_{j=1}^p \log w_{t^j} - \log C( \bs A, \sigma,
\bs m, \bs w) .
\]
This formulation leads to the following sufficient statistics:
\[
S=\Biggl(S_0=\sum_{j=1}^p \mathbh{1}_{t_j=k}, S_1=\sum
_{j=1}^p \mathbh{1}_{t_j=k}b^j\beta^j, \bss\beta\bss\beta
^T, \bs X\bss\beta^T\Biggr).
\]
%

The estimator associated to averaged values of these statistics
(denoted as above with brackets) is
\[
\cases{
\bs A = [\bs X\bss\be^T]([\bss\be\bss\be^T])^{-1} ,\vspace*{2pt}\cr
\sigma^2 = [|\bs X - \bs A\bss\be|^2]/d,\vspace*{2pt}\cr
m_k= [S_1]/[S_0] ,\vspace*{2pt}\cr
w_k= [S_0]/p .}
\]
%

For this model, it is also possible to compute the conditional
distribution of the hidden variables, $\bss\be, \bs t$ and $\bs b$
given observed values of $X$ [\citet{Attias99}]. Indeed, for given
$\bs
b$ and $\bs t$, let $\mu_{\bs b, \bs t} = (b^1m_{t^1}, \ldots, b^p
m_{t^p})$. Let $\Lambda= (\mathrm{Id}_{\mathbb R^p} +
\frac{A^TA}{\sigma^2})$ and, for a given $\mathbf{X}$,
$
\mu_{\bs b, \bs t, \mathbf{X}} = \Lambda(A^T\mathbf{X} +
\mu_{\bs b, \bs t}).
$
Then, a rewriting of the likelihood above shows that the conditional
distribution of $\bss\be$ given $\mathbf{X}, \bs T$ and~$\bs b$
is Gaussian
with mean $\mu_{\bs b, \bs t, X}$ and covariance $\Lambda$, and that
the conditional distribution of $(\bs t, \bs b)$ is
\[
\pi(\bs t, \bs b|X) \propto\exp\biggl(-\frac12\bigl(|\mu_{\bs b, \bs
t}|^2 -
(A^TX + \mu_{\bs b, \bs t})^T \Lambda(A^TX + \mu_{\bs b, \bs
t})\bigr)\biggr)\prod_{j=1}^p w_{t^j}.
\]
Using these expressions, the $E$-step of the EM-algorithm can be
computed exactly, but it requires computing all $(2K+1)^p$ conditional
probabilities~$\pi(\bs t,\allowbreak \bs b|X)$, which becomes intractable for large
dimensions. In contrast, each step of the SAEM algorithm only
requires sampling from the conditional distributions, and has
complexity of order $p(2K+1)$.\vadjust{\goodbreak}

The same remark on the feasibility of the EM algorithm
holds for all our models with discrete variables (BG-ICA,
ET-ICA, etc.), for which the $E$-step of the algorithm can be made explicit
by conditioning on the discrete variables, with a cost that grows
exponentially in the number of components, whereas the sampling part
of SAEM only
grows linearly.

\subsection{BG-ICA and EBG-ICA models}
These two models have the same parameters and maximize the same function.
The likelihood is
\[
-\frac12 \sum_{j=1}^p (Y^j)^2 + \log\biggl(\frac{\alpha}{1-\alpha
}\biggr) \sum_{j=1}^p b^j - \frac{1}{2\sigma^2} \Biggl|X - \sum_{j=1}^p
b^j Y^j \bs a_j\Biggr|^2 - \log C(\sig^2, \bs A, \mu, \alpha)
\]
with sufficient statistic $\bs S\,{=}\,(\bss\be\bss\be^T, \bs X
\bss{\beta}^T, \nu)$ with $\be^j\,{=}\,b^j Y^j$ and \mbox{$\nu\,{=}\,b^1\,{+}\,\cdots\,{+}\,b^p$}.
The optimal parameters
are
\[
\cases{
\bs A = [\bs X\bss\be^T]([\bss\be\bss\be^T])^{-1} ,\vspace*{2pt}\cr
\sigma^2 = [|\bs X - \bs A\bss\be|^2]/d,\vspace*{2pt}\cr
\alpha= [\nu]/p .}
\]

\subsection{ET-ICA, TE-ICA and TEoff-ICA models}

We turn to the ternary models which share the same parameters (up to
$\mu_0$ for the offset model). The likelihood to maximize is
\[
\log\biggl(\frac{\ga}{1-\ga}\biggr) \sum_{j=1}^d |Y^j| -
\frac{1}{2\sigma^2} \Biggl|X - \sum_{j=1}^p s^j Y^j \bs a_j\Biggr|^2 -
\log C(\sig^2, \bs A, \ga)
\]
with sufficient statistic $\bs S = (\bss\be\bss\be^T, \bs X
\bss{\beta}^T, \zeta)$, $\be^j = s^j Y^j$, $\zeta= |Y^1|+\cdots+
|Y^p|$. The optimal parameters are
\[
\cases{
\bs A = [\bs X\bss\be^T]([\bss\be\bss\be^T])^{-1},\vspace*{2pt}\cr
\sigma^2 = [|\bs X - \bs A\bss\be|^2]/d ,\vspace*{2pt}\cr
\ga= [\zeta]/p.}
\]
The maximum likelihood estimator for the single scale model is given by
the same formulae, using $\be^j = sY^j$.
\end{appendix}


\begin{supplement}
\stitle{Supplement to ``A stochastic algorithm for probabilistic
independent component analysis''}
\slink[doi]{10.1214/11-AOAS499SUPP}
\slink[url]{http://lib.stat.cmu.edu/aoas/499/supplement.pdf}
\sdatatype{.pdf}
\sdescription{This file presents a larger version of
some of the images contained in this paper.}
\end{supplement}

%

\printaddresses

\end{document}